\documentclass[aps,twocolumn,showpacs,preprintnumbers,amsmath,amssymb,prb,superscriptaddress]{revtex4-1}

\usepackage{graphicx}
\usepackage{units}
\usepackage{dcolumn}
\usepackage{amsmath,amssymb}
\usepackage{verbatim} 
\usepackage[usenames,dvipsnames]{color}
\usepackage{bm}

\usepackage[tight]{subfigure}
\usepackage{bm}
\usepackage[normalem]{ulem}
\usepackage{enumerate}
\usepackage{mathrsfs,amsmath}

\renewcommand{\vec}[1]{\mathbf{#1}}
\renewcommand{\figurename}{\textbf{Figure}}

\newcommand{\supplabel}{S}

\usepackage[normalem]{ulem} % editing tools: \uuline{urgent}, \uwave{boat}, \sout{wrong},
\usepackage{xr}
\usepackage[a4paper]{hyperref} \hypersetup{colorlinks=true,linktoc=all,linkcolor=Blue,breaklinks=true,citecolor=Blue}
\usepackage[toc,page]{appendix}

\DeclareRobustCommand{\SkipTocEntry}[5]{}

\begin{document}

\title{Size quantization of Dirac fermions in graphene constrictions}

\author{B. Terr\'es}
\affiliation{JARA-FIT and 2nd Institute of Physics, RWTH Aachen University, 52056 Aachen, Germany, EU}
\affiliation{Peter Gr{\"u}nberg Institute (PGI-9), Forschungszentrum J{\"u}lich, 52425 J{\"u}lich,  Germany, EU}

\author{L. A. Chizhova}
\affiliation{Institute for Theoretical Physics, Vienna University of Technology, 1040 Vienna, Austria, EU}
\author{F. Libisch}
\affiliation{Institute for Theoretical Physics, Vienna University of Technology, 1040 Vienna, Austria, EU}

\author{J. Peiro}
\affiliation{JARA-FIT and 2nd Institute of Physics, RWTH Aachen University, 52056 Aachen, Germany, EU}

\author{D. J\"orger}
\affiliation{JARA-FIT and 2nd Institute of Physics, RWTH Aachen University, 52056 Aachen, Germany, EU}

\author{S. Engels}
\affiliation{JARA-FIT and 2nd Institute of Physics, RWTH Aachen University, 52056 Aachen, Germany, EU}
\affiliation{Peter Gr{\"u}nberg Institute (PGI-9), Forschungszentrum J{\"u}lich, 52425 J{\"u}lich,  Germany, EU}

\author{A. Girschik}
\affiliation{Institute for Theoretical Physics, Vienna University of Technology, 1040 Vienna, Austria, EU}

\author{K. Watanabe}
\affiliation{National Institute for Materials Science, 1-1 Namiki, Tsukuba 305-0044, Japan}

\author{T. Taniguchi}
\affiliation{National Institute for Materials Science, 1-1 Namiki, Tsukuba 305-0044, Japan}

\author{S. V. Rotkin}
\affiliation{JARA-FIT and 2nd Institute of Physics, RWTH Aachen University, 52056 Aachen, Germany, EU}
\affiliation{Department of Physics and Center for Advanced Materials and Nanotechnology,
Lehigh University, Bethlehem, Pennsylvania 18015, USA}

\author{J. Burgd\"orfer}
\affiliation{Institute for Theoretical Physics, Vienna University of Technology, 1040 Vienna, Austria, EU}
\affiliation{Institute of Nuclear Research of the Hungarian Academy of Sciences (ATOMKI), 4001 Debrecen, Hungary, EU }

\author{C. Stampfer}
\affiliation{JARA-FIT and 2nd Institute of Physics, RWTH Aachen University, 52056 Aachen, Germany, EU}
\affiliation{Peter Gr{\"u}nberg Institute (PGI-9), Forschungszentrum J{\"u}lich, 52425 J{\"u}lich,  Germany, EU}
\maketitle

%
%  M-x count-words: 149 right now -- Abstract: 150 words max
\textbf{ 
Quantum point contacts (QPCs) are cornerstones of mesoscopic physics and central building blocks for quantum electronics. 
Although the Fermi wavelength in high-quality 
bulk
graphene can be tuned up
to hundreds of nanometers,
 the observation of
quantum confinement of Dirac electrons in 
%low-dimensional
nanostructured graphene systems has proven surprisingly challenging. Here we
show ballistic transport and quantized conductance of size-confined
Dirac fermions in lithographically-defined graphene constrictions. At high charge carrier densities, the observed conductance
agrees excellently with the Landauer theory of ballistic transport
without any adjustable parameter. Experimental data and simulations
for the evolution of the conductance with magnetic field unambiguously
confirm the identification of size quantization in the
constriction. 
Close to the charge neutrality point, 
%Bias
bias voltage spectroscopy reveals a 
renormalized Fermi velocity ($v_{\mathrm F}\approx 1.5\times
10^6$ m/s) in our graphene constrictions.
%, possible related to electron-electron interaction . % [As we discussed not
% ideal, since we do not otherwise discuss e-e interaction in the entire manuscript]
Moreover, at low carrier density transport
measurements allow probing the density of localized states at edges,
thus offering a unique handle on edge physics in graphene devices.  }

The observation of unique transport phenomena in graphene, such as
Klein tunneling~\cite{You09}, evanescent wave transport~\cite{Two06},
or the
half-integer~\cite{Nov05,Zha05} and fractional~\cite{Du09,Bol09}
quantum Hall effect are directly related to the material quality as well as the relativistic dispersion of the charge carriers.  As the quality of
bulk graphene has been impressively improved in the last
years~\cite{Dean10,Wan13}, the understanding of the role and
limitations of edges on transport properties of graphene is becoming
increasingly important. This is particularly true for nanoscale
graphene systems where edges can dominate device properties. Indeed,
the rough edges of graphene nanodevices are most probably responsible
for the difficulties in observing clear confinement induced quantization
effects, such as quantized conductance~\cite{Lin08} and shell
filling~\cite{wang11}.  So far signatures of quantized conductance
have only been observed in suspended graphene, however with limited
control and information on geometry and constriction
width~\cite{Tom11}.  More generally, with further progress in
fabrication technology, graphene nanoribbons and constrictions are
expected to evolve from a disorder
dominated~\cite{Sarma11,Ter11,Danneau08,Borunda13}
%
%[incl. more references?] 
%
transport behavior to a
quasi-ballistic regime where boundary effects, crystal alignment, and
edge defects~\cite{Mas12, Baringhaus14}  govern the
transport characteristics.  
This will open the door to investigate interesting
phenomena arising from edge states, including magnetic order at
zig-zag edges \cite{Magda14}, an unusual Josephson effect,
unconventional edge states
\cite{Plotnik14}, magnetic edge-state excitons~\cite{Yang08} or
topologically protected quantum spin Hall states\cite{Young14}.

%---

%Manifestation of
%Dirac fermions in mesoscopic graphene systems, where the sample
%dimensions become comparable to the mean free path, have led to Klein
%tunneling~\cite{You09}, evanescent wave transport
%phenomena~\cite{Two06} or sub-Poissonian shot noise
%effects~\cite{Dan08}. Unlike these many indications of the
%relativistic nature of charge carriers in ballistic graphene devices,
%observing clear and reproducible confinement quantization effects,
%i.e. quantized conductance, has proven surprisingly
%challenging~\cite{Lin08,Tom11}. With further progress in the
%fabrication methods, graphene is expected to evolve from a disorder
%dominated~\cite{Ter11} transport behavior to a quasi-ballistic regime
%where boundary effects, crystal alignment and edge
%defects~\cite{Mas12} should play a dominant role governing the
%transport characteristics.  Recent advances in the fabrication of
%high-mobility graphene devices~\cite{Wan13} may therefore allow the
%study of confined Dirac fermions in well-defined quasi-one dimensional
%structures.

%-------------------------------------------------------------------------
\begin{figure*}[!t]
  \includegraphics[width=0.98\linewidth]{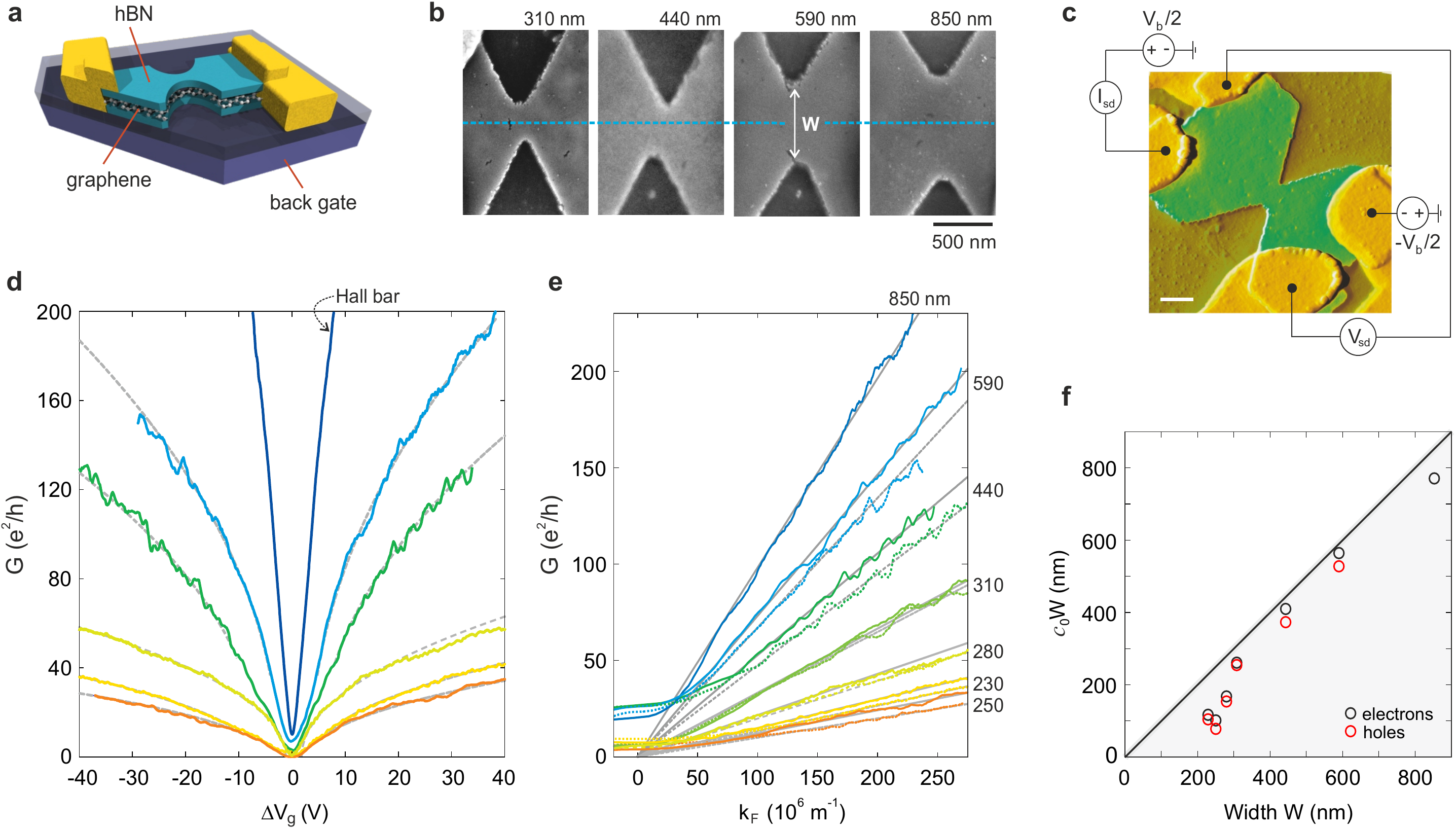}
  \caption{
    \textbf{Width dependent ballistic transport in etched graphene
      nanoconstrictions encapsulated in hBN.}
    \textbf{a}, Schematic illustration of a hBN-graphene sandwich device with the bottom- and top-layers of hBN appearing in green, the gold contacts in yellow, the SiO$_2$ in dark blue and the Si back gate in purple. 
	    \textbf{b}, Scanning electron microscope (SEM) images of four investigated graphene constrictions patterned using reactive ion etching. %The scale bar represents 500 nm. 
    \textbf{c}, False colored atomic force microscope (AFM) image of a fabricated device. Transport is measured in a four-probe configuration to
eliminate any unwanted resistance of the one-dimensional
contacts~\cite{Wan13}. The yellow color denotes the gold contacts, green the top layer of hBN and brown the SiO$_2$ substrate. The white scale bar represents 500 nm. 
    \textbf{d}, Low-bias back-gate characteristics of a Hall bar device (see arrow) and of five constriction devices with different widths ranging from 850 to 230 nm (color code as in panel e). The dashed grey lines are fits to the data. 
		 \textbf{e}, Low-bias four-terminal
    conductance of graphene quantum point contacts as function of
    $k_F$ extracted in the high carrier density limit for seven
    different samples. The color encodes the 
		different samples with different constriction widths (see labels). Grey lines represent a linear fit at high
    values of $k_F$, inserted as guide to the
    eye. Conductance deviates from the expected linear slope for small
    $k_F$. Electron (hole) transport is plotted as solid (dashed) line. Data are taken at temperatures below 2~K.
\textbf{f}, Comparison of $c_0 W$ from conductance traces
(panel e) with the width $W$ (extracted from SEM images).
         }  
  \label{fig:1}
\end{figure*}
%-------------------------------------------------------------------------

%-------------------------------------------------------------------------
\begin{figure*}[!t]
\includegraphics[width=0.95\linewidth]{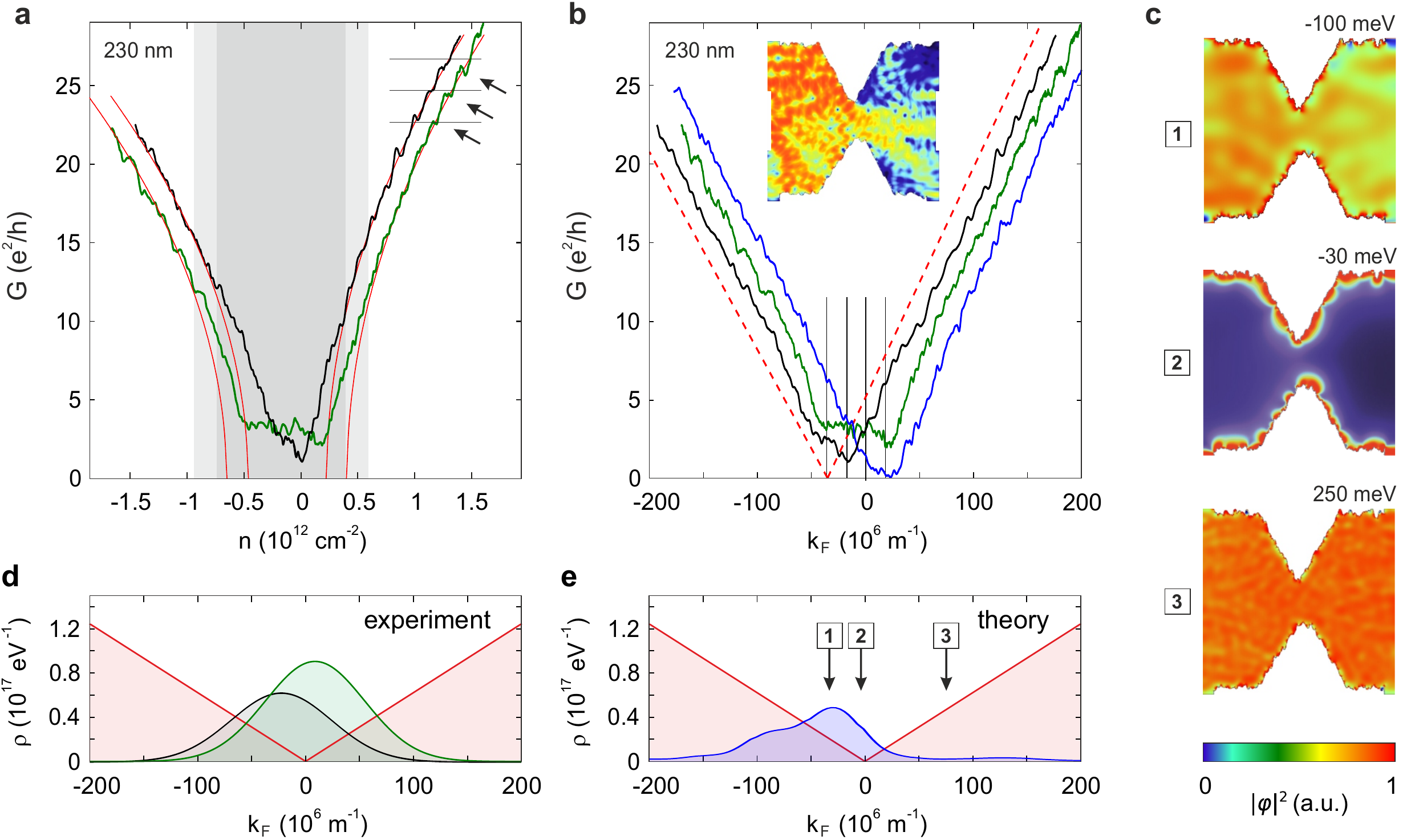}
\caption{\textbf{Conductance through graphene quantum point
    contacts } \textbf{a}, Conductance traces of two different
  cool-downs (black and green curve) of 
the same 
constriction
  ($W \approx 230$~nm) as a function of charge carrier density. For the black
  (green) cool-down, shaded gray (light gray) regions denote
  deviations from the ideal Landauer 
	model
  $G\propto\sqrt{n}$ shown in red. At higher conductance values we observe well reproduced 'kinks' with spacings on the order of 2$e^2/h$ (see arrows and horizontal lines).
\textbf{b}, Experimental conductance trace as a function of $k_F$
after correction for the density of trap states (black and green
curves) and theoretical simulations of graphene quantum point contact
(blue curve). Theoretical results are rescaled to experimental device
size as determined from panel a. Ideal transmission $\propto k_F$ is shown in
red as guide to the eye.  Curves are offset horizontally for clarity.
\textbf{c}, Local density of states of graphene quantum point contact
from tight-binding simulations, at three different energies (-100~meV, 
-30~meV and 250~meV; see also arrows in panel e).
\textbf{d}, Graphene density of states extracted from experiment (fit
to a Gaussian) and \textbf{e} from simulation. Both
experiment and theory find a substantial contribution from trap states
around the Dirac point.}
  \label{fig:2}
\end{figure*}
%-------------------------------------------------------------------------

%-------------------------------------------------------------------------
\begin{figure*}[!t]
  \includegraphics[width=0.75\linewidth]{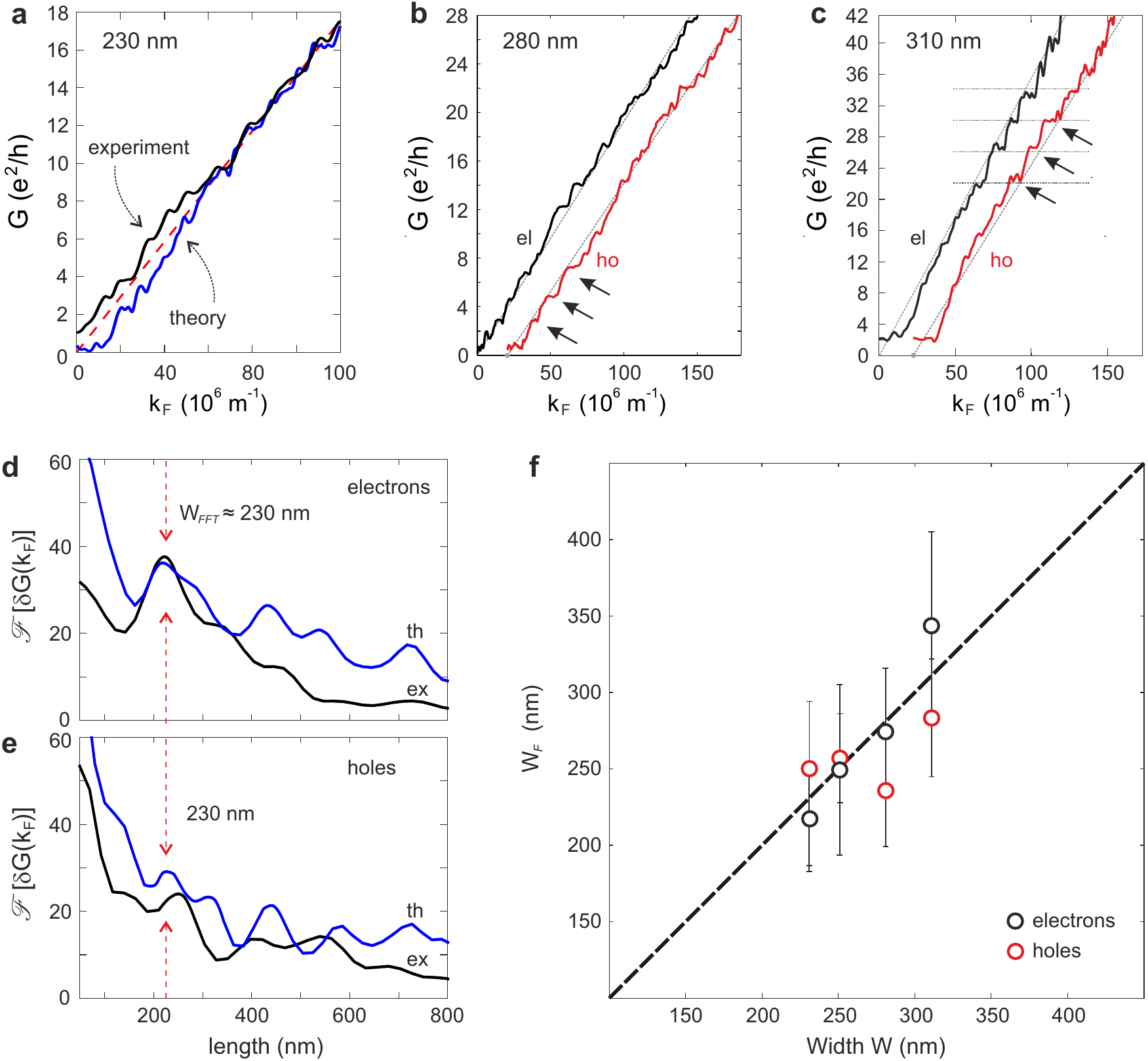}
  \caption{
    \textbf{Size quantization signatures.}
\textbf{a}, Comparison of the low energy conductance between theory (blue) and experiment (black). 
\textbf{b}, \textbf{c}, Measured electron (el - black trace) and hole (ho - red trace) conductance including kink or step-like structure (see arrows) as a function of $k_F$ for two different constriction geometries (see insets). The hole conductance traces are horizontally offest for clarity.
\textbf{d}, Fourier transform of the $G-G^{(0)}$ electron conductance $\mathcal{F}[\delta G(k_{\mathrm{F}})]$ through the 230~nm graphene constriction, for experiment (ex - black trace) and theory (th - blue trace). The first peak of the Fourier transform clearly corresponds to the width $W$ of the quantum point contact (marked by arrows).
\textbf{e}, Same as \textbf{d} for the hole conductance. The size of the first peak is substantially reduced for both experiment and theory due to the presence of localized states that lead to additional scattering.
\textbf{f}, Comparison of width $W_{F}$ extracted from the Fourier transform of
the conductance traces (as in panels e,~f) to geometric constriction width $W$ 
from four different devices (extracted from SEM images).}
\label{fig:3}
\end{figure*}
%-------------------------------------------------------------------------

%-------------------------------------------------------------------------
\begin{figure}[!t]
  \includegraphics[width=1\linewidth]{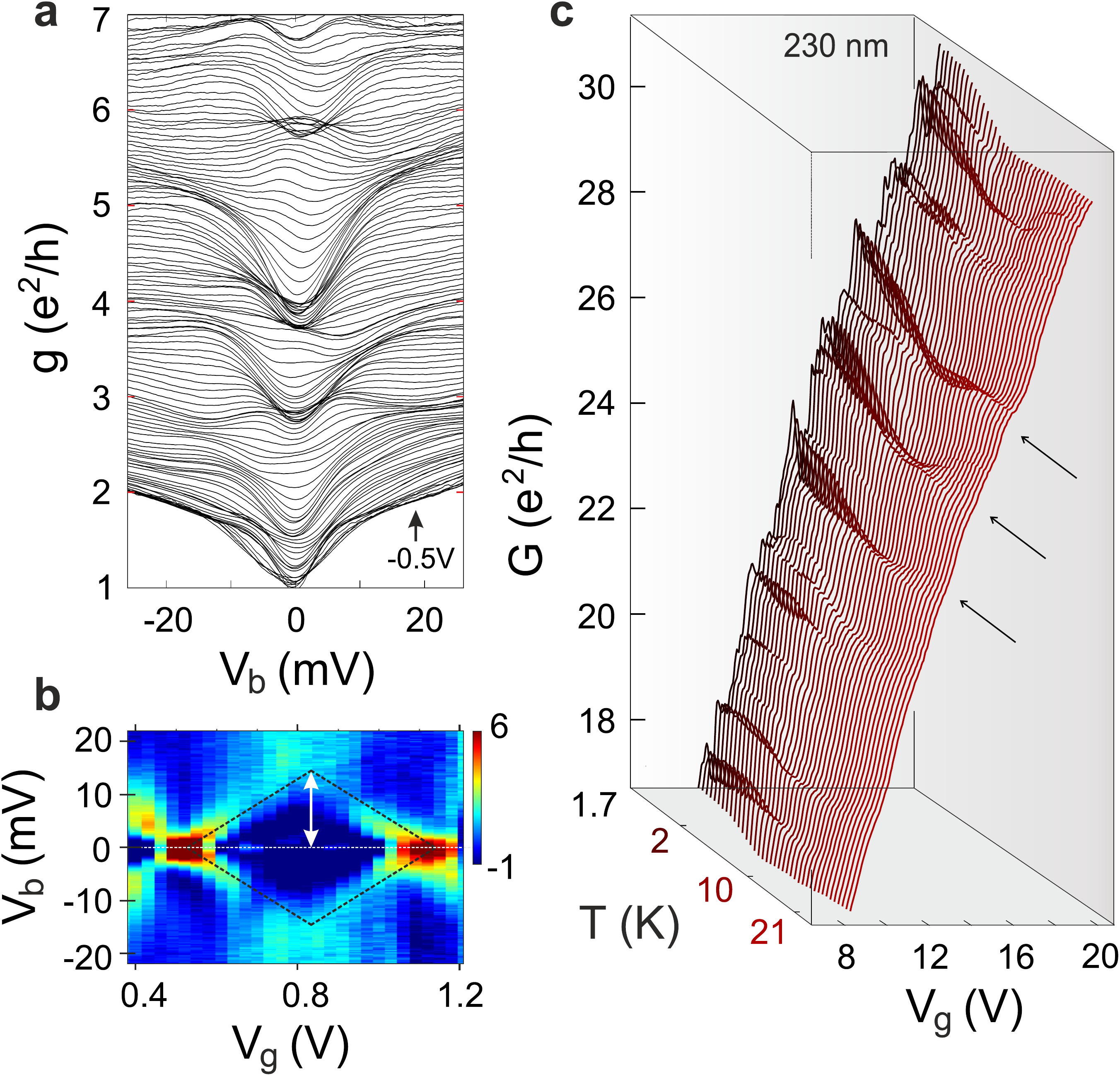}
  \caption{
    \textbf{Quantized conductance: finite bias and temperature dependence.}
\textbf{a}, Zero $B$ field differential conductance $g$ as a function 
of bias voltage $V_{b}$, measured at $T=6$~K, taken at fixed values of 
back-gate voltage $V_{g}$ from $-0.5$~V to $3.0$~V in steps of $30$~mV (see lower right label). 
The dense regions correspond to plateaus in conductance.  
\textbf{b}, Transconductance $\partial  g/\partial V_g$ in units of $e^2/hV$ (see color-scale) as a function 
of bias and back gate voltage for a different cool-down of the same device 
(see also Supplementary Note~6). At $V_{b} = 0$, the transitions between conductance 
plateaus appear as red spots. At finite bias voltage, we observe a diamond 
like shape, which provides an energy scale for the subband energy spacing 
$\Delta E \approx$ 13.5 $\pm$ 2~meV (see dashed black lines and white arrow), which 
is also in good agreement with the energy scale observed in panel a (see also Supplementary Note~6).
\textbf{c}, Conductance traces as a function of temperature and back gate 
voltage. We observe features with different temperature dependencies. Above 
around 10~K only kinks related to quantized conductance survive (see arrows).
 }
  \label{fig:4}
\end{figure}
%-------------------------------------------------------------------------

%-------------------------------------------------------------------------
\begin{figure*}[!t]
  \includegraphics[width=0.98\linewidth]{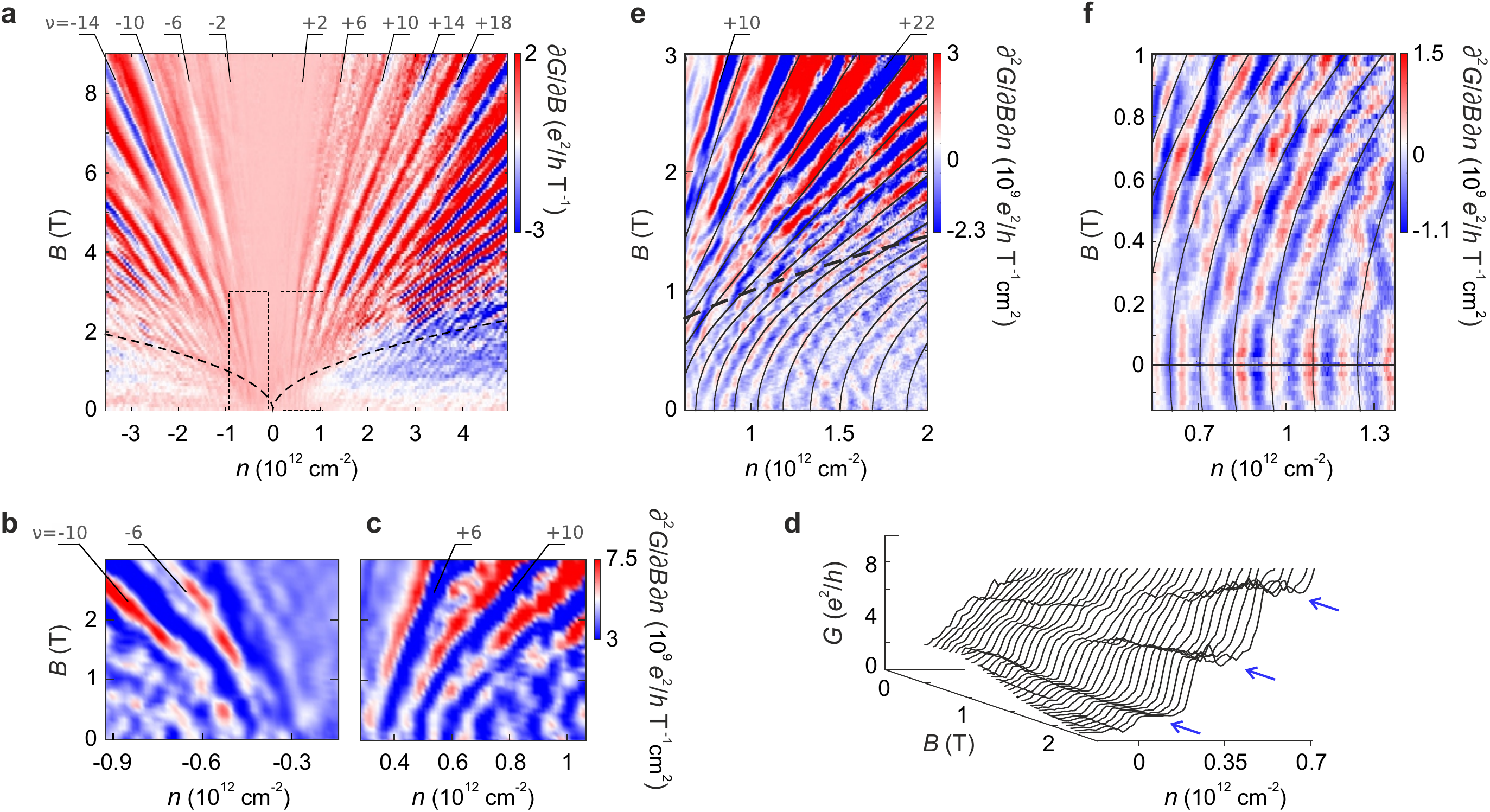}
   \caption{
    \textbf{Magnetic field dependence of the size quantization.}
		\textbf{a}, Landau level fan of the graphene quantum point contact
		 of width $W=230\,nm$, measured at $T=1.7$~K. Landau levels emerge
		at high magnetic fields. The magnetic field quantization of Landau 
		level $m$ dominates over size quantization as soon as $2\sqrt{2m}\,l_B $ 
		(where the magnetic length $l_B \approx 25/\sqrt{B[T]}$ nm) is smaller than 
		the constriction width ($B$ field values above dashed black line). 
  \textbf{b,c}, Double derivative plots of the regions delimited by thin
  dashed lines in panel a showing the evolution of the lowest quantization 
  plateaus with magnetic field: we observe the full transition from quantized 
  sub-bands ($B=0$ T) to Landau levels at large $B$ field.
  \textbf{d}, The same magnetic field evolution is visible in the conductance as a
  function of magnetic field and charge carrier density for a different cool-down 
 of the same device, also measured at $1.7\,K$.
The blue arrows highlight the expected quantum Hall conductance plateaus at 
2, 6 and 10~$e^2/h$. 
 \textbf{e}, Double derivative plot of the conductance as a function of
  magnetic field and charge carrier density measured at $T=6\,K$. The solid 
	black lines denote the theoretical expectations for the evolution of the size 
	quantization with magnetic field. The thick dashed black line corresponds to the 
	boundary of the Landau level regime, also appearing in panel a.
 \textbf{f}, Zoom-in of panel e for small magnetic fields $B$ $\le 1$ T.
	}
	\label{fig:5}
\end{figure*}
%-------------------------------------------------------------------------

In this work we report on the observation of quantum confinement and edge states in
ballistic transport through graphene constrictions approximating quantum point contacts.
We prepared 4-probe devices based on 
high-mobility graphene-hexagonal boron nitride (hBN)
sandwiches on SiO$_2$/Si substrates and use reactive ion etching to pattern
narrow constrictions (see Methods) with widths ranging from $W\approx$~230 to 850~nm, connecting wide leads (Figs.~1a-1c). 
The graphene leads are side-contacted~\cite{Wan13} by %80~nm thick 
chrome/gold electrodes. A back gate voltage is applied on the highly doped Si
substrate to tune the carrier density in the graphene layer, $n=\alpha
(V_g - V_g^0)=\alpha \Delta V_g$, where $\alpha$ is the so-called
lever arm and $V_g^0$ is the gate voltage of the minimum conductance,
i.e. the charge neutrality point. 
% Tuning the gate voltage $V_g$
%thus allows changing the Fermi wavevector $k_F = \sqrt{n\pi}$ of conducting electrons on the Dirac cone.
To demonstrate the high electronic quality of our graphene-hBN sandwich
structures we show the gate characteristic of a reference Hall bar
device (Fig.~1d).
From this data we extract a carrier mobility in the range of around 150.000~cm$^2$/Vs
(see Supplementary Note~1), resulting in a mean free path exceeding 1~$\mu$m at around $\Delta V_g = 4.6$~V. Thus, the mean free path is expected to clearly exceed all relevant length scales in
our constriction devices giving rise to ballistic transport.

\addtocontents{toc}{\SkipTocEntry}
\section*{Results}
\addtocontents{toc}{\SkipTocEntry}
\subsection*{Ballistic transport.}
We measure the conductance as function of gate voltage 
for a number of constrictions with different widths $W$ 
(Fig.~1d; see labels in Fig.~1e). The observed square 
root dependence $G \propto \sqrt{\Delta V_g} \propto \sqrt{n}$
(see dashed lines in Fig.~1d) is a first indication of 
highly ballistic transport in our devices.
Indeed, according to the Landauer theory for ballistic 
transport, the conductance through a perfect constriction 
increases by an additional conductance quantum $e^2/h$ 
whenever $W k_{\mathrm{F}}$ reaches a multiple of $\pi$,
\begin{equation}
%G = \frac{4e^2}{h} \left[ \frac{1}{2} + \sum_{m=1}^\infty \theta\left(\frac{W k_{\mathrm{F}}}{\pi}-m\right) \right],
G = \frac{4e^2}{h} \sum_{m=1}^\infty \theta\left(\frac{W k_{\mathrm{F}}}{\pi}-m\right),
\end{equation} 
where $k_F = \sqrt{\pi n}$ is the Fermi wave number, 
the factor four accounts for the valley and spin 
degeneracies, $\theta$ is the step function, and we 
have neglected minor phase contributions due to details 
of the graphene edge\cite{Ostaay11} for simplicity. 
%%NEW
%Assuming an additional energy-independent average transmission coefficient $\overline{T}$ and
%taking the 
Fourier expansion of Eq.~(1) yields
\begin{equation}
G = \frac{4e^2}{h}\frac{c_0 W k_{\mathrm{F}}}{\pi}+\frac{4e^2}{h}
  \left[\sum_{j=1}^\infty c_j\sin\left(2 j W k_{\mathrm{F}} - \phi_j\right) - \frac{c_0}{2}\right].
%	G = \frac{4e^2}{h}\frac{\overline{T} W k_{\mathrm{F}}}{\pi}-\frac{4e^2}{h}
%  \left[\frac{1}{2} - \sum_{j=1}^\infty a_j\sin\left(2 j W k_{\mathrm{F}}\right) \right],
%\approx \frac{4e^2}{h}\frac{W}{\sqrt\pi} \sqrt{\alpha \Delta V_g},
% \frac{4e^2}{h} \frac{W}{\pi} k_F 
\end{equation}
For an ideal constriction $c_0$ = 1, $\phi_j = 0$, and 
$c_j = 1/(j\pi)$, $j > 0$.  In the presence of edge 
roughness, $c_0$ is reduced to a value below 1 due to 
limited average transmission, and higher Fourier components 
$c_j$ are expected to decay in magnitude and acquire 
random scattering phases $\phi_j \ne 0$.  Consequently,
the sharp quantization steps turn into periodic modulations 
as will be shown below.  Averaged over these modulations 
only the zeroth order term in the expansion [Eq.~(2)] 
survives.  This mean conductance $G^{(0)}$ of a constriction 
of width $W$ thus features a linear dependence on 
$k_{\mathrm{F}}$, or, equivalently, a square-root dependence 
as a function of back-gate voltage assuming an energy-independent 
transmission $c_0$ of all modes, in accord with Fig.~1d.
%, $G^{(0)}\approx 4e^2W/h\cdot
%\sqrt{\alpha \Delta V_g/\pi}$,
%
By measuring the carrier density dependent quantum 
Hall effect at high magnetic fields\cite{YZhang05,Novoselov07}, 
we can independently determine the gate coupling 
$\alpha$ for each device (see Supplementary Note 2). We 
can thus unfold the dependence on $V_g$ and study both 
the electron and hole conductance as function of 
$k_{\mathrm{F}}$ (Fig.~1e). From the linear slopes of 
$G(k_{\mathrm{F}})$, the product $c_0 W$ can be extracted 
for each device and compared to its width $W$ (Fig.~1f) 
determined from scanning electron microscopy (SEM) images 
(see, e.g., Fig.~1b).  The estimates for $c_0 W$ extracted
from $G^{(0)}$ lie just little below the width $W$, where $c_0$ 
decreases for decreasing width.
This suggests that for the narrower devices reflections, 
most likely due to device geometry and edge roughness, are 
playing a more important role. From the data in Fig.~1f 
we can extract $c_0 \approx 0.56$ for our smallest constriction.
Below we will show that, indeed, reflections at the rough edges 
of the constriction and not a reduction in active channel width 
is responsible for the deviation of the experimentally extracted 
$c_0 W$ from the SEM width $W$. 

%the assumption of reflection-less transmission
%(Eq.~2) is not valid
%where the rouh edges
%of our constrictions result in an increased partial reflection of the
%charge carriers.  As we will show below, the average transmission
%coefficient can be estimated by $W/W_g$, leading e.g. to a value of
%$\approx 0.56$ for our smallest constriction.

\addtocontents{toc}{\SkipTocEntry}
\subsection*{Localized states at the edges.}

For small $k_F < 50 \times 10^6$~m$^{-1}$ (i.e. low carrier
concentrations) the measured conductances systematically deviate 
from the expected linear behavior (see Fig.~1e). 
This deviation from the square-root relation between $G$ and $n$
(i.e. $\Delta V_g$) becomes more apparent when focusing on $G$ around
the charge neutrality point (CNP). The conductance as function
of $n$ for two different cool-downs of the same graphene constriction
($W \approx$~230~nm, Fig.~2a), shows marked cool-down dependent low
carrier density regions with substantial deviations from $G \propto
\sqrt{n}$. Far away from the CNP, the conductance as function of 
$n$ for both cool-downs shows (i) an identical $\sqrt{n}$ behavior 
leading to the very same $c_0 W$ and (ii) almost identical, regularly 
spaced kink structures (see arrows in Fig.~2a), which are, however, 
slightly shifted relative to another on the carrier density axis $n$. 
These observations suggest that the square-root relation between the 
Fermi wave vector $k_F$ and the gate voltage $V_g$, i.e. $n$ needs to 
be modified.
%: it assumes constant gate coupling and
%the ideal linear density of states $\rho \propto k_{\mathrm F}$ of
%Dirac fermions. 
While the quantum capacitance of ideal graphene can 
be neglected \cite{Ilani06, Fang07,Reiter14}, a small additional
contribution $n_T(\Delta V_g)$ from, e.g., localized trap states
modifies the relation between $n$ and $k_F$ to
\begin{equation}
\alpha \Delta V_g = n = k_F^2\pi^{-1} + n_T\left(\Delta V_g\right).
\end{equation}
Far away from the Dirac point ($k_{\mathrm{F}}^2 \gg \pi n_T$), we
recover the expected square root relation. Close to the Dirac point,
however, $\alpha \Delta V_g$ will be strongly modified by deviations
$n_T$ from the linear density of states of ideal Dirac fermions and
approaches $n_T(\Delta V_g)$ near the CNP.

The trap states do not 
contribute to transport, yet they contribute to the charging 
characteristics~\cite{Bischoff14}. It is
important to note that electron-hole puddles~\cite{LeRoySTM} or
charged impurities would only smear out the density of states but
would not add additional trap-state density $n_T$. 
This is in contrast to graphene edges, in particular rough 
graphene edges, which feature a significant number of trap states.
For example, a tight-binding
simulation of the local density of states of the experimental
geometry yields a strong clustering of localized states at the
device edges (see Fig.~2c), which energetically lie close to the CNP
(Fig.~2e). The deviation of $G$ from the $\sqrt{n}$ scaling also opens 
up the opportunity to extract $n_T$ from experimental conductance data 
(e.g. Fig.~2d), and thus a new pathway for device characterization. 
Inspired by the tight-binding simulation, we approximate the trap 
state density $n_T$ as function of Fermi wave vector by a Gaussian 
distribution. We fit the position, height and width of the Gaussian 
by minimizing the difference between the measured $G(k_{\mathrm{F}})$ 
and the corresponding linear extrapolation to very low values of 
$k_{\mathrm{F}}$ (see Fig.~2b and Supplementary Note 3).
%We fit $n_T(k_{\mathrm{F}})$ to a Gaussian to preserve
%  the small-scale oscillations of the measured conductance.
%
We find good qualitative agreement between simulation and
experiment (compare Figs.~2d and 2e).  Quantitative correspondence
would require a detailed, microscopic model for the trap state
density $n_T$. Note that the only difference between different
traces in Figs.~2a, 2b and 2d is the exposition of the device to air
for several days leading to a wider carrier density region of
substantial deviations (green trace). The number of trap states
(i.e., the deviations around the CNP) is significantly enhanced
(compare also green and black trace in Fig.~2d). As the active
graphene layer is completely sandwiched in hBN only the graphene 
edges are exposed to air and, very likely, experience chemical 
modifications. In line with our numerical results, we thus 
conjecture that localized states at the edges substantially 
contribute to $n_T$ , leading to the strong cool-down dependence 
we observe in our measurements. While this interpretation seems 
plausible and is consistent with our data, alternative explanations 
cannot be ruled out.

%\todo{ Joachim would strongly advise to remove this sentence! 
%Further investigations, however, are necessary for unraveling 
%the microscopic nature of $n_T$.}
%Interestingly, also the minimum conductance
%appears to be cool-down dependent (compare minima of black and green
%traces in Figs.~2a, 2b).

Away from the CNP our data agrees remarkably well with ballistic 
transport simulations through the device geometry using a modular 
Green's function approach \cite{LibischNJP} (see blue trace in 
Fig.~2b): we simulate the 4-probe constriction geometry taken 
from a SEM image, scaled down by a factor of four to obtain a 
numerically feasible problem size~\cite{liu15}. To account for 
the etched edges in the devices, we include an edge roughness 
amplitude of $\Delta W = 0.2 W$ for the constriction. This comparatively 
large edge roughness (which is consistent with the systematic 
reduction of transmission through the constriction when using
the average conductance) is probably due to microcracks at the 
edges of the device.
%
%Using the extracted
%density of states, we can finally obtain a conductance as linear
%function of Fermi energy (Fig.~2e).
%%%%%%%%%%%%%
%%%%%%%%%%%%%

\addtocontents{toc}{\SkipTocEntry}
\subsection*{Quantized conductance.}

Superimposed on the overall linear behavior of 
$G(k_{\mathrm{F}})$, we find reproducible modulations 
(``kinks'') in the conductance (see Figs.~3a-3c and 
Fig.~S4b). The kinks are well reproduced for several 
cool downs (see arrows in Fig.~2a and Supplementary 
Note 4) as well as for different devices, generally 
showing a spacing $\Delta G$ varying in the range of
$(2 - 4) e^2/h$ (see arrows in Figs.~3b and 3c). 
The ``step height'' and its sharpness depend on the 
carrier density (i.e. $k_F$) as well as on the constriction 
width and is strongly influenced by the overall transmission 
$c_0$ (Fig.~1f). Remarkably, we observe a spacing 
$\Delta G$ of the steps close to $4 e^2/h$ for
one of our wide samples ($W$ $\approx$ 310~nm) at elevated 
conductance values on both the electron and hole sides 
(see arrows and horizontal lines in Fig.~3c and Fig.~S4b)

Our assignment of the conductance ``kinks'' as signatures 
of quantized flow through the constriction is supported 
by our theoretical results. Theory and experimental data 
from the smallest constriction show similar smoothed, 
irregular modulations (see Fig.~3a), instead of sharp size 
quantization steps.\cite{Per06} The replacement of sharp 
quantization steps by kinks reflects the strong scattering 
at the rough edges of the device~\cite{Muc09,Ihnatsenka12},
resulting in the accumulation of random phases in the Fourier
components 
of $G$ [Eq.~(2)]. We note that calculations with smaller
edge disorder show a larger average conductance, yet very 
similar ``kink'' structures. As the present calculation 
includes only edge-disorder induced scattering while 
neglecting other scattering channels such as electron-electron 
or electron-phonon scattering, the good agreement with 
the data suggests edge scattering to be the dominant 
contribution to the formation of the ``kinks''. By contrast,
both experimental and theoretical investigations of, e.g.,
semiconducting GaAs heterostructures show very clear, 
pronounced quantization plateaus\cite{vanWeesConductance}. 
In these heterostructures, the electron wave length near 
the $\Gamma$ point is very long, and cannot resolve edge disorder on the nanometer scale. By
contrast, $K$-$K'$ scattering in graphene allows conduction electrons
to probe disorder on a much shorter length scale. 
%\new{This argument
%implies broken valley symmetry, consistent with the observed step height
%of only $2 e^2/h$ (since physical spin should be conserved).} 
Consequently, edge
roughness substantially impacts transport.  The comparison between
experimental and theoretical data (Fig.~3a) unambiguously establishes
the observed modulations to be consistent with the smoothed size
quantization effects predicted by theory.

By subtracting the zeroth-order Fourier component $\propto
k_{\mathrm{F}}$ (or $\sqrt n$), the superimposed modulations 
of the conductance $\delta G(k_{\mathrm{F}}) = G - G^{(0)}$ 
provide direct information on the quantized conductance through 
the constriction [Eq.~(2)]. One key observation is that the 
Fourier transform of $\delta G(k_{\mathrm{F}})$ offers an 
alternative route towards the determination of the constriction 
width complementary to that from the mean conductance 
$G^{(0)}$. For example, the pronounced peak of the first 
harmonic at $230 $ nm (red arrows in Figs.~3d and 3e) is 
consistent with the constriction width $W$ derived from
the SEM image.  Interestingly, our simulation also correctly
reproduces the experimental observation that the peak in the 
Fourier spectrum of $\delta G(k_{\mathrm{F}})$ is more 
pronounced on the electron side (Fig.~3d) than on the hole 
side. This results from the slightly asymmetric energy 
distribution of the trap states relative to the CNP, which 
is accounted for in our tight-binding calculation.

Performing such a Fourier analysis for several devices (Supplementary 
Note 5) yields much closer agreement with the geometric width $W$ 
(Fig.~3f and horizontal axis of Fig. 1f) than an estimate based 
only on the zeroth-order Fourier component $c_0 W$ [first term in 
Eq.~(2), see vertical axis of Fig.~1f]. Fourier spectroscopy of 
conductance modulations thus allows to disentangle reduced transmission 
due to scattering at the edges ($\tilde c_0 W$) from the effective 
width of the constriction, and proves the relation between the 
observed Fourier periodicity and the device geometry.
%Fourier spectroscopy of conductance modulations thus allows to
%pinpoint edge scattering as reason for the reduced transmission.

Bias voltage spectroscopy measurements yield an estimate for 
the energy scale of the size quantization steps~\cite{Tom11,wep13}.
For example, by analyzing finite bias measurements from our
smallest constriction device we extract a subband energy 
spacing of $\Delta E =$~13.5 $\pm$ 2~meV near the CNP (Figs.~4a,~4b 
and Supplementary Note~6).
%$\Delta E \approx$~14~meV (Figs.~4a,~4b).  
With the geometric width of 230~nm also confirmed by the
Fourier spectroscopy (Fig. 3c) we can
estimate the Fermi velocity near the CNP as $v_F = 2 W \Delta E/h
= (1.5 \pm 0.2) \times 10^6$~m/s. This is a clear signature of a
substantially renormalized Fermi velocity in nanostructured graphene,
possibly enhanced by electron-electron interaction~\cite{Elias11}.
Moreover, the extracted energy scales are consistent with the weak temperature
dependence of the quantized conductance (Fig.~4c and Supplementary Note~7).

\addtocontents{toc}{\SkipTocEntry}
\subsection*{Transition from quantized conductance to quantum Hall.}
Additional clear fingerprints of size quantization appear in the
parametric evolution of the conductance steps~\cite{gui12} with
magnetic field, $B$. The transition from size quantization at zero
$B$-field to Landau quantization at high magnetic fields occurs when
the cyclotron radius $l_C$ becomes smaller than half the constriction
width $W$.  For the Landau level $m$ the transition should occur at $2
\, l_C = 2\,\sqrt{2m}\,l_B \approx W$ with $l_B$ the magnetic
length. This transition line in the $B-n$ plane (see black dashed
curve in Fig. 5a) agrees well with the onset of Landau level formation
in our data (see Supplementary Note~8 for similar data from a 280~nm constriction device). The evolution of the lowest quantized
steps (at $B$ = 0 T) to the corresponding lowest Landau levels at low
temperatures (T=1.7 K) can be easily tracked (see Figs.~5b and~5c).
At higher temperatures ($T = 6$ K) the evolution of quantized
sub-bands to Landau levels is observed even for higher conductance
plateaus (Fig.~5d,~5e). For a comparison, we calculate the evolution
of size quantization of an infinitely long ribbon of width $W$ as
function of magnetic field. We take $W \approx 230$ nm from the SEM
data, which leaves no adjustable parameters. Our model ( black lines
in Figs.~5e and 5f) reproduces the evolution from the kinks at small
fields ($l_B\gg W$) to the Landau levels for large fields ($l_B < W$)
remarkably well, further supporting the notion that they are, indeed,
signature of size quantization.

\addtocontents{toc}{\SkipTocEntry}
\section*{Discussion}

%In summary, 
We have shown ballistic conductance of confined
Dirac fermions in high-mobility graphene nanoconstrictions sandwiched
by hexagonal boron nitride. Away from the Dirac point, we observe a
linear increase in conductance as function of Fermi wavevector with a
slope proportional to constriction width. Close to the Dirac point,
the charging of localized edge states distorts this linear relation.
Superimposed on the linear conductance, we observe reproducible,
evenly spaced modulations (``kinks''). Tight-binding simulations for
the device reproduce these structures related to size quantization at
the constriction.  
%We can unambiguously identify the evolution of size quanti-
%zation steps with magnetic field 
We can unambiguously identify these ``kinks'' as size quantization
signatures by both Fourier spectroscopy at zero magnetic field and
their evolution with magnetic field, finding good agreement between
theory and experiment.

%================================================================================================
%  Reduce fontsize for the rest % {fontsize}{spacingsize}
\begingroup \fontsize{8}{9}\selectfont
%

%================================================================================================
\addtocontents{toc}{\SkipTocEntry}
\section*{Methods}
\textbf{Experimental methods and details}

The hBN-graphene-hBN sandwich structures~\cite{Wan13} have been etched 
by reactive ion etching in a $SF_6$ atmosphere, prior deposition of a 
$\sim\!10\,nm$-thick Cr etching mask. Remaining rests of Cr oxide are 
removed by immersing the samples in a Tetramethylammonium hydroxide 
(TMAH) solution for about 30-35 s. All transport measurements are 
performed in a 4-probe configuration using standard lock-in techniques.
Since the distances between the contacted current-carrying electrodes 
and the voltage probes are small, compared to the other length scales of the
system, we have an effective 2-probe configuration. Importantly,
this way we exclude the one-dimensional contact resistances. %For
%further details, please see the \suppinfo.
% with an
%applied excitation voltage of $V_{AC}=250\,\mu V_{RMS}$.

\textbf{Electrostatic simulations and transport calculations}

We simulate the experimental device geometry using a third-nearest
neighbor tight-binding ansatz. We rescale our device by a factor of
four compared to experiment, to arrive at a numerically feasible
geometry. We determine the Green's function using the modular
recursive Green's function method \cite{Rotter06, LibischNJP}.
The local density of states and transport properties can then be
 extracted by suitable projections on the Green's function.
For more technical details see Supplementary Note~9.

%\textbf{Parameter estimation}
%%
%%For this analysis it is crucial that measurements with vibration couplings switched ``off'' are available, providing access to the ``bare'' parameters, see \methods.
%%
%????.
%
\addtocontents{toc}{\SkipTocEntry}
\bibliographystyle{naturemag}

%\bibliography{cite}

%\addtocontents{toc}{\SkipTocEntry}

%\vspace{-0.5cm}
%================================================================================================
\addtocontents{toc}{\SkipTocEntry}
\acknowledgments
We acknowledge stimulating discussions with F. Hassler, F. Haupt and B.J van Wees.
Support by
the HNF, the DFG (SPP-1459), the ERC (GA-Nr. 280140), the EU project Graphene
Flagship (Contract No. NECT-ICT-604391) and Spinograph, and
the Austrian Science Fund (SFB-041 VICOM and DK-W1243 Solids4Fun) is gratefully
acknowledged.
%We acknowledge the financial support from the DFG under Contract No.~SPP-1243, and
%the Austrian Science Fund, SFB-041 VICOM. 
Calculations were performed on
the Vienna Scientific Clusters.

Correspondence and requests for materials should be addressed to F.L. and C.~S.

%================================================================================================
\vspace{-0.5cm}
\addtocontents{toc}{\SkipTocEntry}
\subsection*{Author contributions}
B.T. and C.S. conceived the project.
B.T. fabricated the samples, performed the experiments and interpreted the data.
S.E. assisted during measurements.
B.T. and D.J analyzed the data.
L.A.C. and F.L. performed the numerical calculations and theoretical analysis, A.G. and F.L. developed the numerical code.
T.T. and K.W. synthesized the hBN crystals.
J.B., S.V.R. and C.S. advised on theory and experiments.
B.T., L.A.C, F.L., J.B., and C.S. prepared the manuscript.
All authors contributed in discussions and writing of the manuscript.
%================================================================================================

\vspace{-0.5cm}
\addtocontents{toc}{\SkipTocEntry}
\subsection*{Competing financial interests}
The authors declare no competing financial interests.

\newpage

\newpage
\endgroup

\pagebreak

%\documentclass[aps,preprint,showpacs,preprintnumbers,amsmath,amssymb,prb,superscriptaddress]{revtex4-1}

%\linespread{1.5}

%\usepackage{graphicx}
%\usepackage{units}
%\usepackage{dcolumn}
%\usepackage{amsmath,amssymb}
%\usepackage{verbatim}
%\usepackage[usenames,dvipsnames]{color}

%\usepackage[tight]{subfigure}
%\usepackage{amsmath}
%\usepackage{bm}
%\usepackage[normalem]{ulem}
%\usepackage{enumerate}
%\usepackage{mathrsfs,amsmath}

%\widowpenalty10000
%\clubpenalty10000

% size of the text space
%\addtolength{\oddsidemargin}{1.25cm}
%\addtolength{\evensidemargin}{1.25cm}
%\addtolength{\textwidth}{-3cm}
%\addtolength{\marginparwidth}{0cm}
%\addtolength{\parindent}{1em}

%\addtolength{\headsep}{0cm}
%\addtolength{\textheight}{0cm}
%\addtolength{\headheight}{0cm}
%\addtolength{\topmargin}{0cm}

%\usepackage[normalem]{ulem}
%
\renewcommand\textfraction{.01}

\renewcommand{\vec}[1]{\mathbf{#1}}
%\newcommand{\vecg}[1]{\pmb{ #1}}
%\newcommand{\veco}[1]{ \hat{ \mathbf{#1}} }
%\newcommand{\tens}[1]{\mathbf{#1}}
%\newcommand{\tenso}[1]{\hat{\mathbf{#1}}}
%\newcommand{\bra}[1]{\langle #1|}
%\newcommand{\ket}[1]{|#1 \rangle}
%\newcommand{\braket}[1]{\langle #1 \rangle}
%\newcommand{\tr}{\text{tr}}
%\newcommand{\Tr}{\text{Tr}}
%\newcommand{\suppinfo}{Supplementary Information }

% Label prefix for suppinfo numbers:
%\newcommand{\supplabel}{S}
%\newcommand{\figopt}{t}

% Referencing suppinfo equations / figures
%\newcommand{\Eq}[1]{equation~(\ref{#1})}
%\newcommand{\eq}[1]{(\ref{#1})}
%\newcommand{\Fig}[1]{Fig.~\ref{#1}}
%\newcommand{\fig}[1]{\ref{#1}}
%\newcommand{\Sec}[1]{Sec.~\ref{#1}}
%\newcommand{\Ref}[1]{Ref.~[\onlinecite{#1}]}  % no brackets for Nat. Phys. !
% NOTE: citations are NOT modified (e.g., to Ref. [S1], technically difficult...)

% Formatting of equation / figure numbers
\renewcommand{\figurename}{\textbf{Supplementary Figure~}}
\renewcommand{\tablename}{\textbf{Supplementary Table~}}
\renewcommand{\thetable}{\supplabel\arabic{table}}
\renewcommand{\thefigure}{\supplabel\arabic{figure}}
\renewcommand{\theequation}{\supplabel\arabic{equation}}
\renewcommand{\thesubsection}{Supplementary Note~\arabic{subsection}}

% Change reference part title
\renewcommand\refname{Supplementary References}

% Referencing external labels (in main paper text)
%\usepackage{xr}
\externaldocument[E]{../paper}

\onecolumngrid

%\newpage
%\posttitle{\begin{center}\large Size quantization of Dirac fermions in graphene constrictions \\ \\
%\bigskip
%\LARGE Supplementary Information
%}
%\maketitle
%\end{center}
%\newpage

%\vspace*{-2em}

%\newpage
%\addtocontents{toc}{\SkipTocEntry}
%\section*{\hfil Size quantization of Dirac fermions in graphene constrictions \quad %\quad \quad \hfil}
%\vspace{42}
%\addtocontents{toc}{\SkipTocEntry}
%\section*{\hfil Supplementary Information \quad \quad \quad \quad \quad \quad \hfil}
%\newpage

\newpage

\addtocontents{toc}{\SkipTocEntry}
\section*{\LARGE Supplementary Information}

%\vspace{42}
%\addtocontents{toc}{\SkipTocEntry}
%\section*{\hfil Supplementary Information \quad \quad \quad \quad \quad \quad \hfil}

\newpage

%\addtocontents{toc}{\protect\sloppy} % fixes long titles running into the margin of TOC !
                                     % requires "breaklinks=true" option for hyperref above !
																	
\tableofcontents
%---------------------------------------------------------------------------------------------

%---------------------------------------------------------------------------------------------
%\begin{abstract}
\vspace*{2em} In this Supplementary Information we provide additional
experimental data as well as a detailed description of the
experimental and theoretical methods expanding the `Methods' section
of the main article.  Within the Supplementary Information, references 
are numbered as, e.g., equation ({\supplabel}1) and Figure {\supplabel}1, 
whereas regular numbers, e.g., equation (1) and Figure 1, refer to the main
article.
%\end{abstract}
%---------------------------------------------------------------------------------------------
\setcounter{figure}{0}
\setcounter{equation}{0}

\newpage

\subsection{Sample quality}

 The field-effect carrier mobility in our sandwich devices is on the
 order of $150.000$ cm$^2$/Vs. This high sample quality is thanks to advances
 in sample fabrication, in particular the van-der-Waals stacking
 process: the graphene is fully
 encapsulated in hBN, resulting in significantly improved sample
 quality. We extract the mobility from a one $\mu$m-wide Hall bar device
 fabricated in the very same batch as our graphene constrictions (see
 Fig.~S1). The dark blue trace in Fig.~1d of the main
 manuscript is taken from this Hall bar device. 
As all traces from the
 constrictions with different widths (some of them carved out from the
 same hBN-graphene-hBN sandwich) lie systematically below the Hall
 bar trace, we exclude bulk scattering as limiting process in our devices.
 Independently, we have shown recently in a collaboration with
 A. Morpurgo, F. Guinea and coworkers~\cite{Couto14} that in our
 high-quality devices the carrier mobility is not limited by charge
 impurity and short-range scattering but rather by nanometer-scale
 strain variations giving rise to long-range scattering with allowed
 pseudospin flips. We expect that the same limitations on the mean
 free path also apply to our graphene constriction devices.

\begin{figure*}[!h]
  \includegraphics[width=0.84\linewidth]{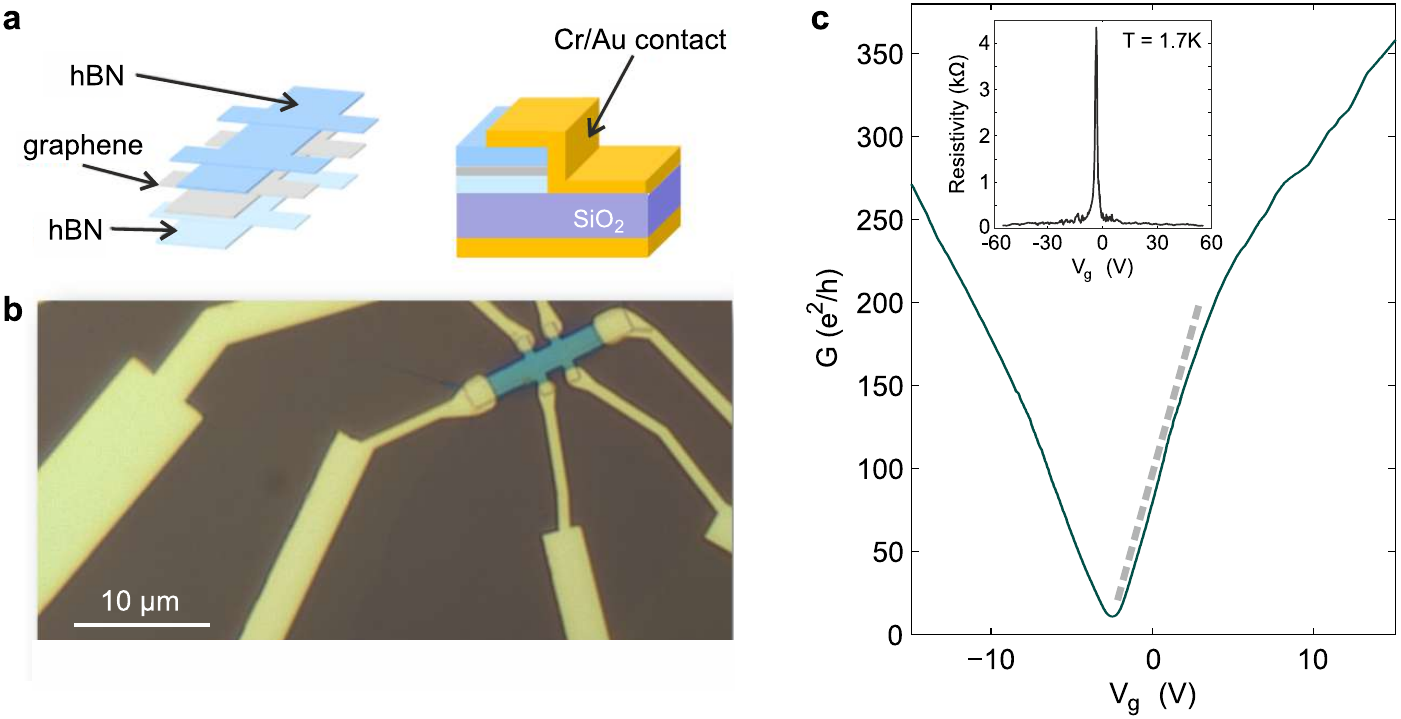}
\caption[Fig.S2]{\textbf{Reference hBN-graphene-hBN sandwich Hall bar device.}
\textbf{(a)} Schematic illustration of the hBN-graphene-hBN sandwich structure and nature of the quasi-one-dimensional graphene-metal (Cr/Au) contact.
\textbf{(b)} Optical image of an etched and contacted $\sim$1~$\mu$m-wide hBN-graphene-hBN sandwich Hall bar device.
\textbf{(c)} Four-terminal conductance as a function of back gate voltage $V_g$ measured at a constant current of 50~nA and a temperature of 16~K. From the linear slope near the charge neutrality point (see dashed line) we extract a carrier mobility of around 150.000~cm$^2$/Vs. The inset shows the four-terminal resistivity as a function of gate voltage at lower temperature (1.7~K).}
	\label{fig:reference hall bar}
\end{figure*}

\newpage

\subsection{Extraction of the gate lever arm $\alpha$}

\begin{figure*}
  \includegraphics[width=0.75\linewidth]{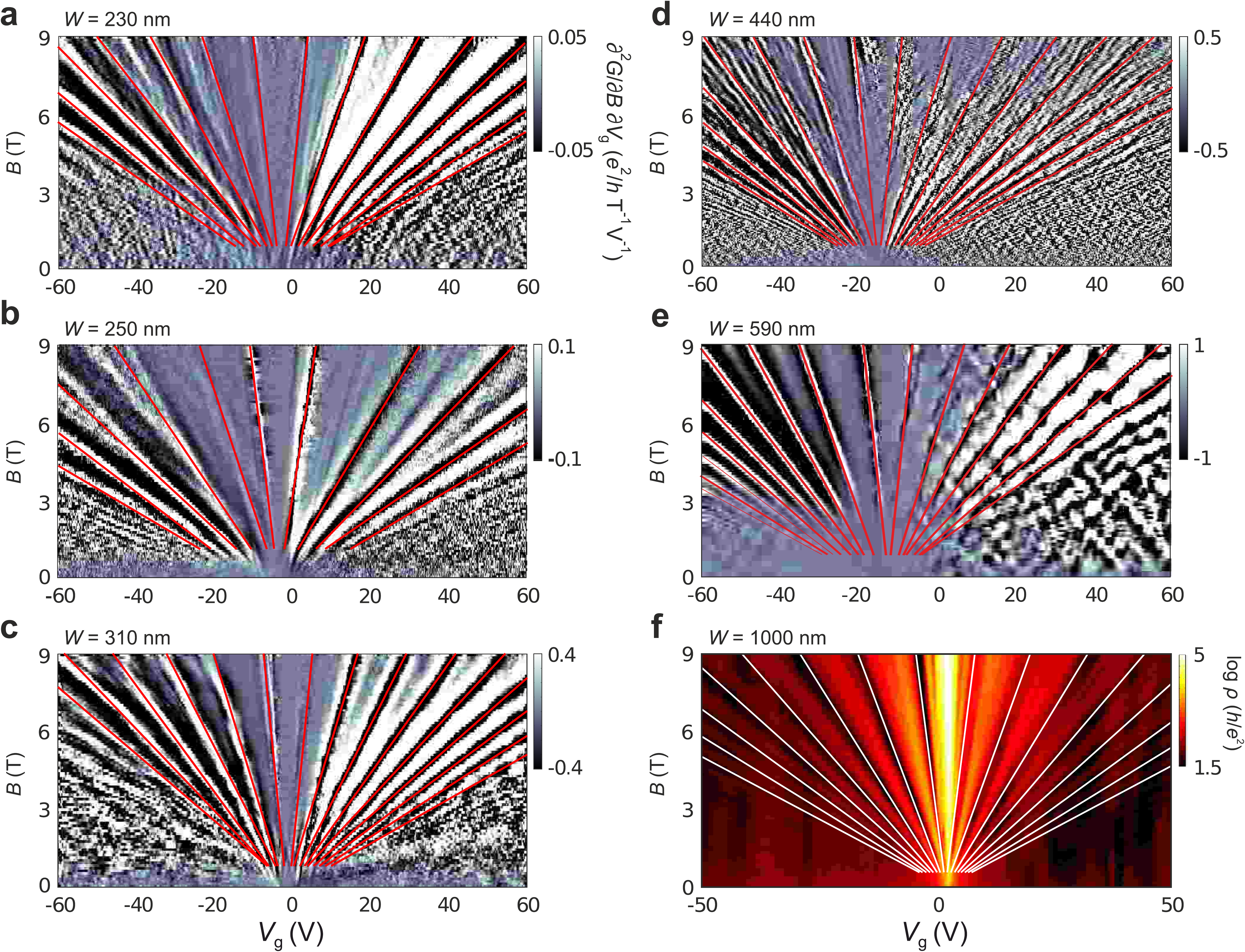}
\caption[Fig.S2]{\textbf{Landau fan and capacitive coupling.} 
\textbf{(a)}-\textbf{(e)} Second derivative of the longitudinal conductance $\partial^2 G/ \partial V_g \partial B$ as a function of magnetic field $B$ and back-gate voltage $V_g$ for six different devices with different widths. The red lines follow the evolution of the Landau levels. The slopes of the lines are proportional to the capacitive coupling $\alpha$. 
\textbf{(f)} The longitudinal resistivity $\rho$ as a function of $B$ and $V_g$ provide an alternative way to extract $\alpha$ from the position of the Landau levels, marked by white lines.
}
	\label{fig:leverarm}
\end{figure*}

Measurements of Landau levels in graphene as a function of back 
gate voltage $V_g$ and magnetic field $B$ (see Fig.~\ref{fig:leverarm}) 
allow for an independent determination of the gate coupling (or lever 
arm) $\alpha$. The Landau level spectrum for massless Dirac fermions in graphene is
given by
\begin{equation}
E_m(B) = \text{sgn}(m) v_F \sqrt{2|e|\hbar |m|B}, \ \ \ m\in \mathbb{Z}_0,
\label{eq:landaulevels}
\end{equation}
where $v_F$ is the Fermi velocity and $m$ is the quantum number 
of the corresponding Landau level. 
Assuming a perfect linear dispersion and a constant capacitive gate 
coupling leads to the following relation between energy $E$ and back gate voltage 

\begin{equation}
E = \hbar v_F k_F = \hbar v_F \sqrt{\pi \alpha \Delta V_g},
\label{eq:E_V}
\end{equation}
where $\Delta V_g = V_g - V^0_g$, and $V^0_g$ is the gate voltage at
the charge neutrality point. As a result, the Landau levels in the $B$
- $V_g$ plane form straight lines, i.e.~$B_m = C_m \Delta V_g$, where
the slope $C_m = \alpha h/4me$ is Landau level index ($m$) dependent and proportional to the capacitive
coupling $\alpha$ (see red lines in Fig.~\ref{fig:leverarm}a-e). %

\begin{table}
\begin{center}
\begin{tabular}{l||lllll}
\hline 
SEM width $W$ (nm)  &  $\alpha$ ($10^{10}/cm^{2}V$)  \\%& $V_{0}\,(V)\,@\,0\,T$ \\%& $V_{0}\,(V)\,@\,9\,T$ & $\Delta V_{0}\,(V)$ \\
\hline
1000  & $7{.}00$  \\% & $-2{.}63$ \\% & - & - \\ % Hall bar
850   & $5{.}80$  \\% & $-1{.}66$ \\% & - & - \\ % E
590   & $6{.}75$    \\%& $-13{.}00$ \\% & - & - \\ % C
440   & $6{.}90$    \\%& $-14{.}85$ \\% & - & - \\ % D
310   & $7{.}00$    \\%& $-1{.}20$  \\% & - & - \\ % A
280   & $7{.}20$    \\%& $-7{.}20$  \\% & - & - \\ % B
250   & $5{.}40$    \\%& $-7{.}20$  \\% & - & - \\ % B
230   & $7{.}15$     \\%& $-2{.}30$  \\% & - & - \\ % ER
\hline
\end{tabular}
\end{center}
\caption{
Lever arm values $\alpha$ for eight different devices extracted from the Landau level fan measurements (see Fig.~\ref{fig:leverarm}). %(R) corresponds to an extended quantum point contact or ribbon.
}
\label{tab:alphas}
\end{table}

The onset of each Landau level can be resolved by taking the mixed
second derivative of the longitudinal conductance $G$ with respect to
$V_g$ and $B$, i.e.~$\partial^2 G/ \partial V_g \partial B$. The
positions of the Landau levels coincide with the minima/maxima of the
derivative on the electron/hole side (see Fig.~\ref{fig:leverarm}a-e,
where the local minima/maxima coincide with red lines). Alternatively,
the Landau levels can be determined from the minima of the
longitudinal resistivity $\rho$ (marked in white in
Fig.~\ref{fig:leverarm}f).  Note that $C_m$ is independent of the
Fermi velocity, experimental determination of which is rather
difficult.  Table~\ref{tab:alphas} summarizes the extracted values of
$\alpha$ for the different devices.

\clearpage
%==========================================================================================
\subsection{Linearization of $G$ as a function of $k_F$}

For a known gate coupling $\alpha$, one can evaluate the
measured conductance $G(V_g)$ as a function of $k_F$, 
using the standard constant capacitive coupling model $k_F=\sqrt{\pi \alpha \Delta V_g}$. 
Following the Landauer theory of conductance 
through a constriction of finite width $W$, the averaged 
conductance $G^{(0)}(V_g)$ features a square-root dependence 
on $V_g$,
\begin{equation}
G^{(0)}=\frac{4 e^2}{h}\left( \frac{c_0 W k_F}{\pi} -\frac{c_0}{2}\right) = \frac{4 e^2}{h}\frac{c_0 W}{\pi}  \sqrt{\pi\alpha (V_g - V_g^0)} -\frac{2 e^2}{h}c_0.
\label{eq:ballistic}
\end{equation}
\begin{figure*}
  \includegraphics[width=0.77\linewidth]{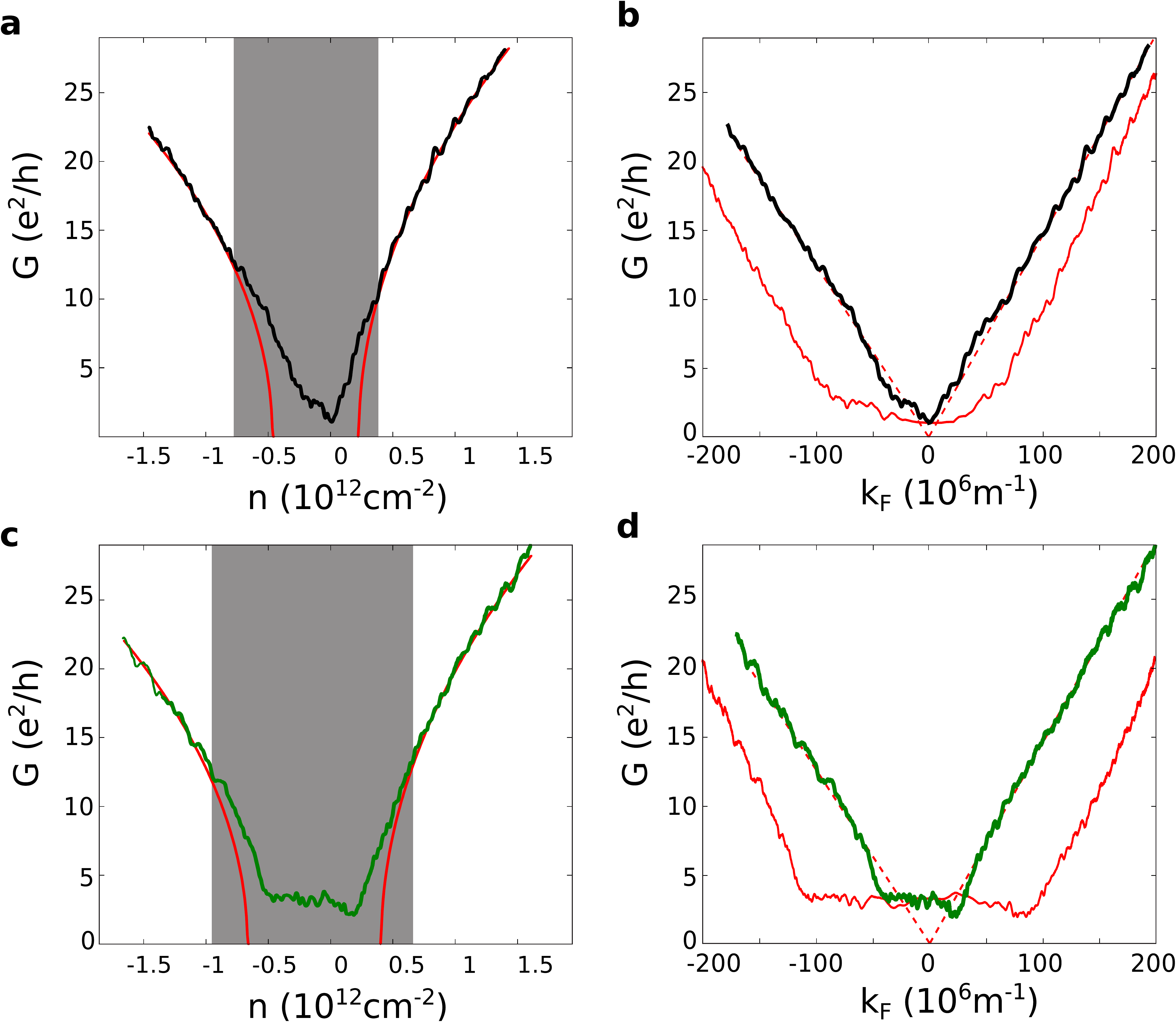}
\caption[Fig.2]{\textbf{Deviation from the ideal ballistic conductance for the 230 nm-wide graphene constrictions.}
\textbf{(a)} Low-bias four-terminal conductance $G$ as a function of
charge carrier density $n$. The red solid lines are fits to a simple capacitive
coupling model [Eq.~(\ref{eq:E_V})] valid at high carrier densities for the
holes and electrons regime, respectively. Deviations appear in the
gray-shaded region around the charge neutrality point.
\textbf{(b)} Conductance $G$ of panel (a) as a function of $k_F$ using
the linear density of states of ideal graphene (red solid line), or
including a finite density of trap states (Eq.~\ref{rhoNT}) around the
Dirac point (black solid line). The linear relation between $G(k_F)$
and $k_F$ expected for ideal graphene is shown as a 
dashed red line.
\textbf{(c)} and \textbf{(d)} Corresponding to (a) and (b) but for a different
cool-down of the same constriction. After exposing the sample to
ambient conditions, the number of charge traps responsible for the
flat area around the Dirac point increased significantly.
}
	\label{fig:Ballistic_Regime}
\end{figure*}

A closer look at the traces from two different cool-downs of the
narrowest device with $W = 230$~nm (Figs.~\ref{fig:Ballistic_Regime}a and \ref{fig:Ballistic_Regime}c)
reveals a systematical deviation from the expected square-root
dependence of $G$ [Eq.~(\ref{eq:ballistic})] at low
carrier concentrations, i.e for $n < 0.45 \times 10^{12}$~cm$^{-2}$ on
the electron side and $n <0.75 \times 10^{12}$~cm$^{-2}$ on the hole
side (Fig.~\ref{fig:Ballistic_Regime}a). This deviation
becomes more pronounced closer to the charge neutrality point (see
shaded area in Figs.~\ref{fig:Ballistic_Regime}a and~\ref{fig:Ballistic_Regime}c). In the ballistic region, i.e., far from the charge neutrality point, we can use
Eq.~(\ref{eq:ballistic}), with $\alpha$ extracted from
the Landau level fan, and fit parameters $V_g^{0,e}$ for the electron
($e$) and $V_g^{0,h}$ for the hole ($h$) side. As expected, the conductance $G$ evolves
linearly as function of $k_F$ in the ballistic regime (see red traces
in Figs.~\ref{fig:Ballistic_Regime}b and \ref{fig:Ballistic_Regime}d),
but large deviations between data and model become apparent close to
the charge neutrality point. We conclude that a linear model using a
constant gate coupling is not directly applicable to our graphene
constriction devices. %The culprit is the assumption of a perfectly
%linear density of states of ideal Dirac fermions implicitly included
%in the capacitive coupling model of Eq.~(\ref{eq:E_V}).
 Instead, one needs to account for the additional charge carrier trap
states $n_T$ (see main text), modifying the relation between back-gate
voltage and Fermi wave number according to 
\begin{equation}
\alpha ( V_g - V_g^0) = \alpha \Delta V_g = k_F^2\pi^{-1} + n_T\left(\Delta V_g\right).\label{rhoNT}
\end{equation}
 Using Eq.~\ref{rhoNT}, we obtain an implicit mapping $k_{\mathrm{F}}(\Delta V_g)$, which
 depends on the functional form of $n_T(\Delta V_g)$ and accounts for the modified 
density of states in the constriction,
\begin{equation}
k_{\mathrm{F}}(\Delta V_g) = \sqrt{\pi\alpha \Delta V_g - \pi n_T\left(\Delta V_g\right)}.
\label{rescaledkf}
\end{equation}
We conjecture that the strong cool-down dependence seen in the
different traces of Fig.~\ref{fig:Ballistic_Regime} are due to
modifications in the trap state densities as the sample was exposed to
air~\cite{Woe16}. As the graphene layer in our
hBN-graphene-hBN sandwich can only interact with air at the edges,
edge states presumably strongly contribute to $n_T$. Indeed,
tight-binding simulations of the constriction geometry (see main text,
  Fig.~2c,e) yield a clustering of localized edge states close to the
Dirac point.  Accounting for $n_T$ by Eq.~(\ref{rescaledkf}) should
recover the linear relation between Fermi wave number and
conductance. We can thus determine $n_T$ from the measured
conductance: we assume a Gaussian distribution of trap states $n_T$,
and fit the width, position and height of the Gaussian
distribution by minimizing deviations of the
rescaled conductance $G[(k_{\mathrm{F}}(\Delta V_g)]$ from the linear
conductance $G^{(0)}(k_{\mathrm{F}})$ of Eq.~(\ref{eq:ballistic}), see green/black traces in
Figs.~\ref{fig:Ballistic_Regime}b and
\ref{fig:Ballistic_Regime}d. Note that this procedure assumes that
any other sources for a deviation from a linear relation between $k_F$ and $G$ (due to,
e.g., many-body effects) are small compared to the contribution from trap states $n_T$.

\subsection{Reproducibility of kink signatures}

\begin{figure*}[b]
    \includegraphics[width=0.82\linewidth]{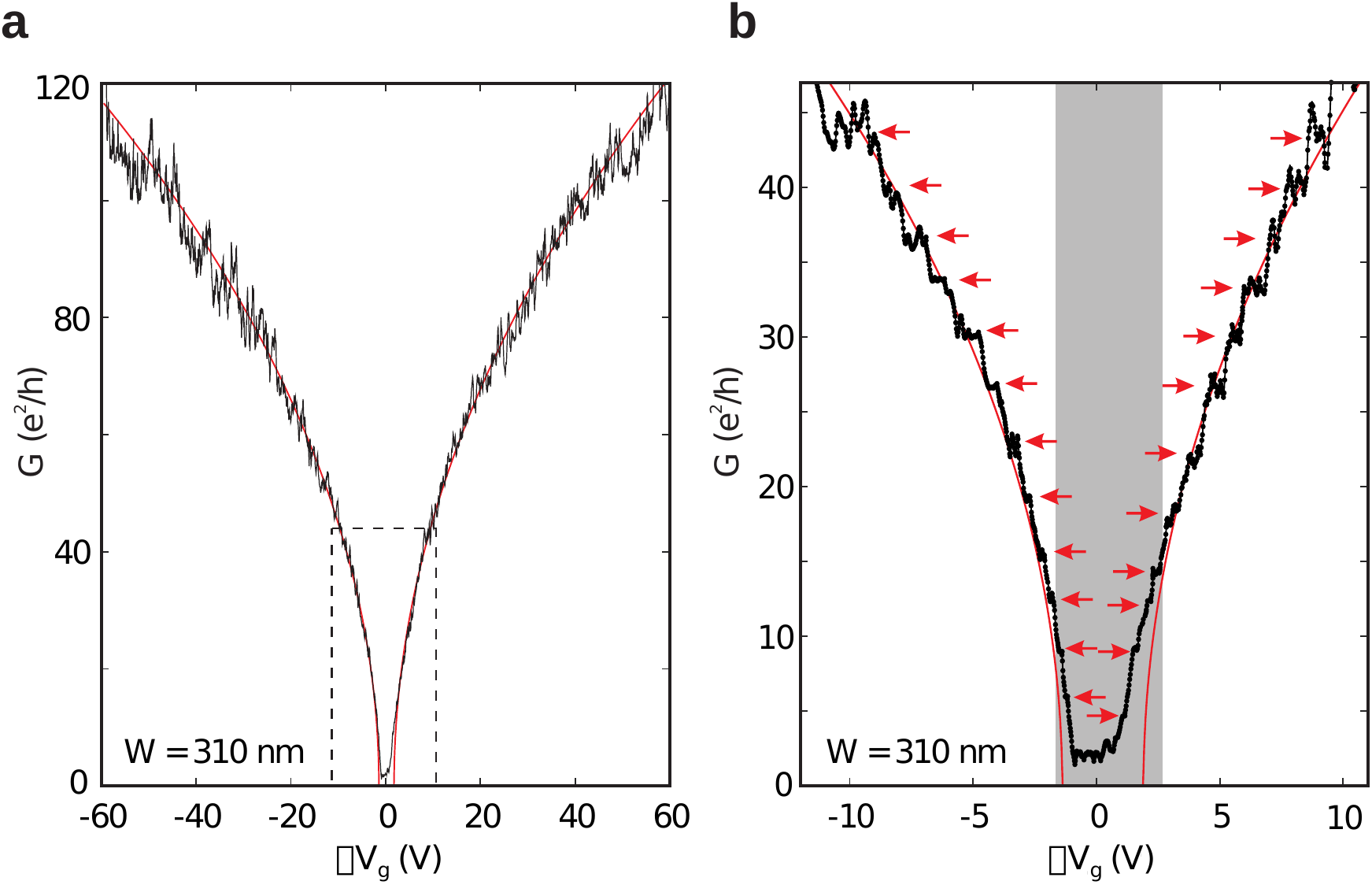}
\caption[Fig.S4]{\textbf{Kinks in the back-gate characteristics of the $310\,$nm-wide graphene constriction}
\textbf{(a)} Low-bias four-terminal conductance $G$ as a function of back gate voltage $V_g$, measured at $T=2$ K. The ideal Landau-B\"uttiker model of conductance $G \propto \sqrt {n}$ is marked in red.
\textbf{(b)} Close-up of the conductance $G$ inside the dashed-line region of panel a. The reproducible kinks are clearly visible (marked by red arrows). The shaded gray region denote deviations from the ideal Landauer model (red trace).
}
  \label{fig:plateaus_310nm}
\end{figure*}

\begin{figure*}[!h]
  \includegraphics[width=0.36\linewidth]{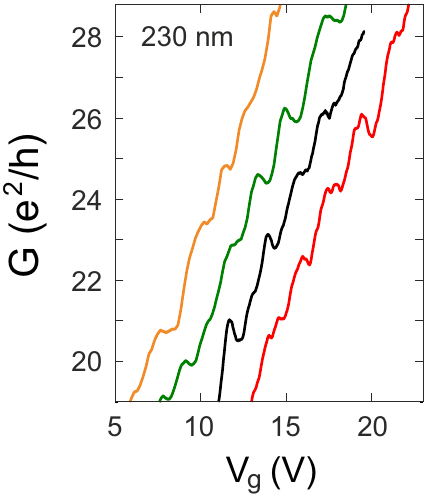}
\caption[Fig.S2]{\textbf{Cool-down dependence of the kinks for the 230 nm-wide graphene constriction I.}
Four-terminal conductance $G$ as a function of back gate voltage $V_g$ for four different cool-downs of the $230\,$nm-wide graphene constriction. The traces are shifted horizontally for clarity.
}
	\label{fig:MultipleCoolDown1}
\end{figure*}

\begin{figure*}[hb]
  \includegraphics[width=0.53\linewidth]{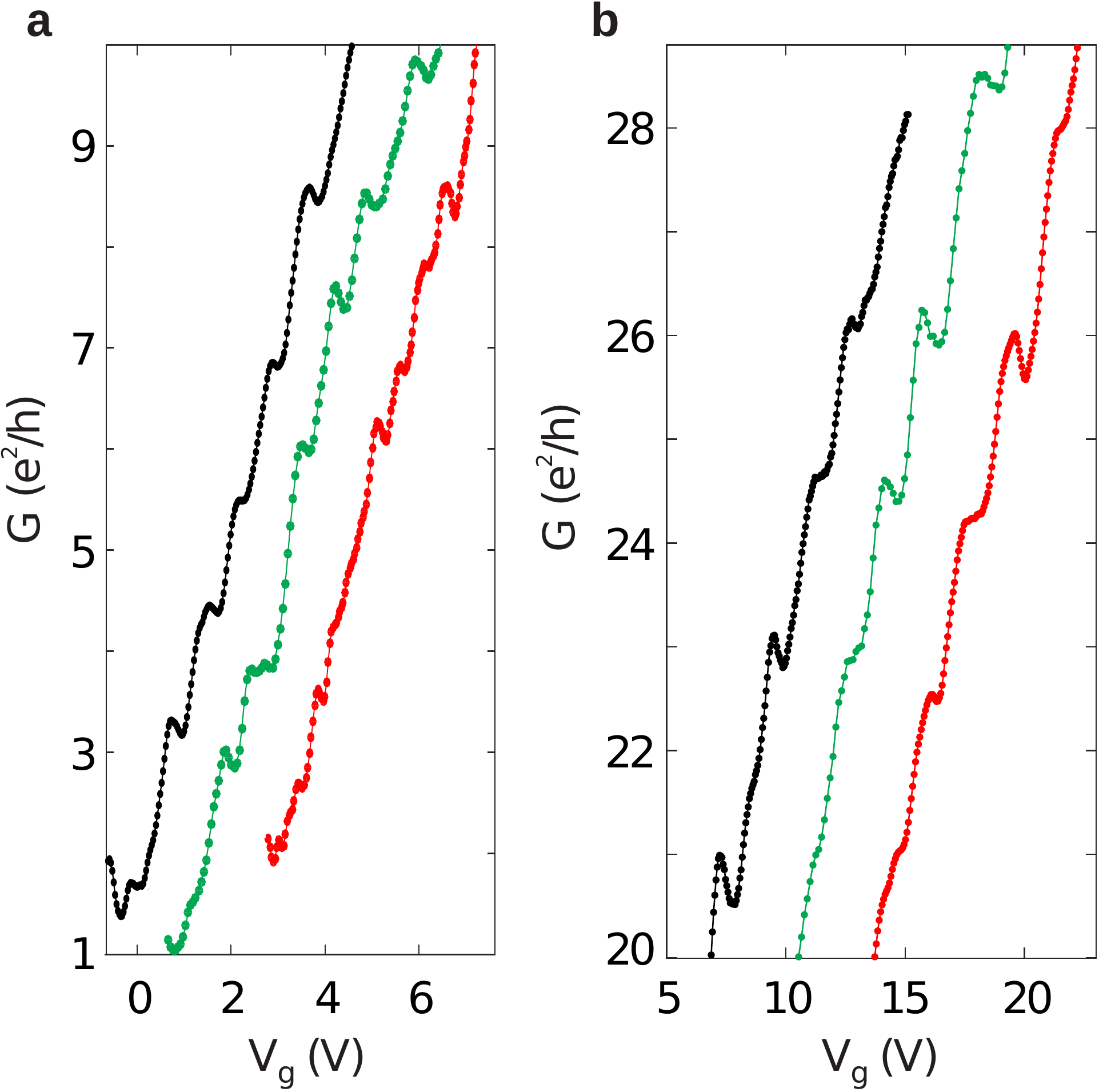}
\caption[Fig.S2]{\textbf{Cool-down dependence of the kinks for the 230 nm-wide graphene constriction II.}
\textbf{(a)} and \textbf{(b)} Four-terminal conductance $G$ as a function of back gate voltage $V_g$ for different cool-downs of the $230\,$nm-wide graphene constriction at low and high charge carrier densities (panels a and b, respectively). The traces are shifted horizontally for clarity.
}
	\label{fig:MultipleCoolDown}
\end{figure*}

\begin{figure}[!b]\centering
\includegraphics[draft=false,keepaspectratio=true,clip,%
                   width=0.65\linewidth]%
                   {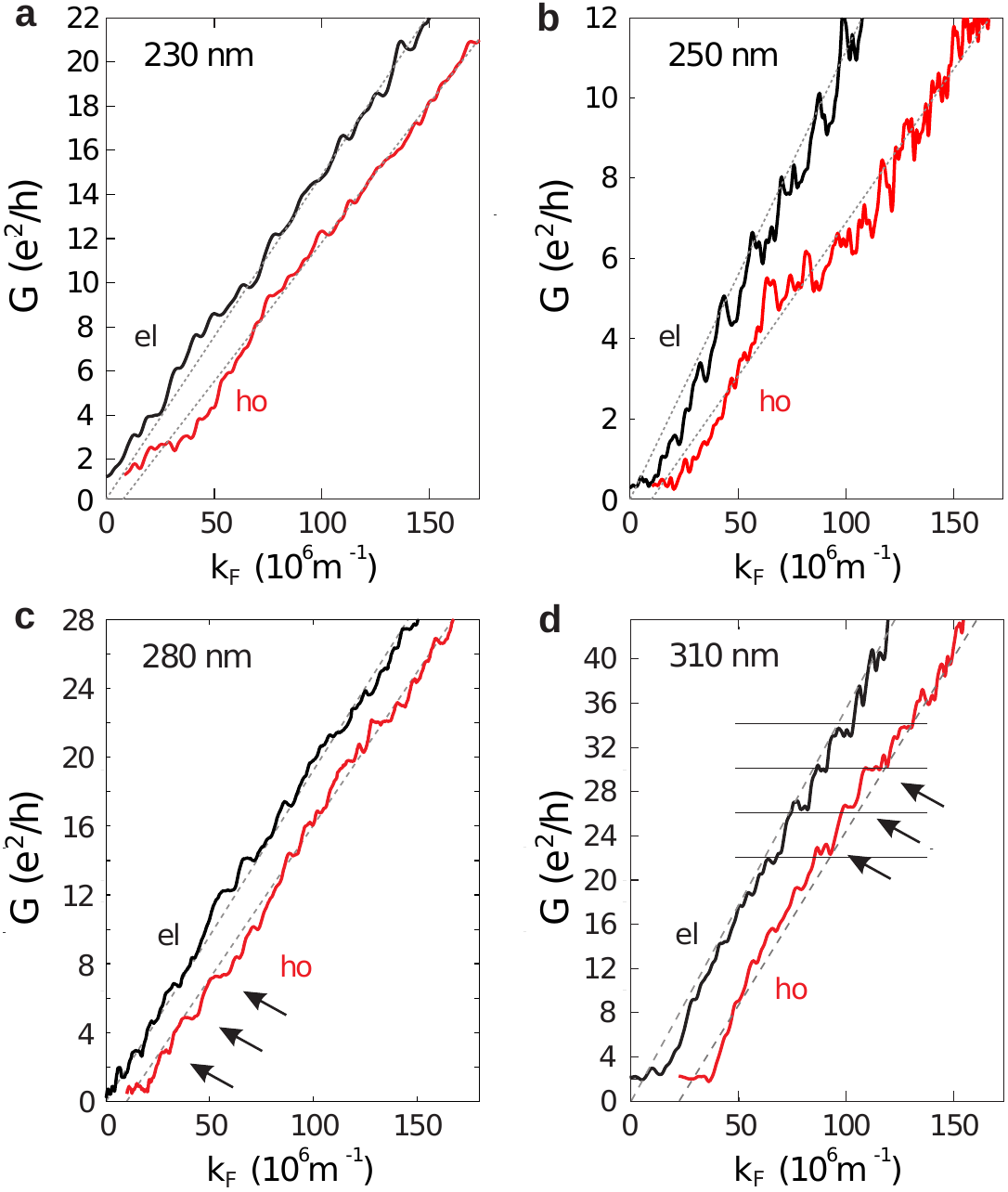}                   
\caption[FIG1]{\textbf{Width dependence of the kinks in conductance.} 
Four-terminal conductance $G$ as a function of back gate voltage $V_g$
for four different devices of widths $230\,$nm (\textbf{a}), $250\,$nm
(\textbf{b}), $280\,$nm (\textbf{c}) and $310\,$nm (\textbf{d}). The
transmission traces are shown in black (red) for electrons (holes) as
a function of rescaled $k_F$ (see main text). The arrows point to
kinks where the conductance jumps by about $c_0\times
\frac{4e^2}{h}$, with $c_0$ as measure for the overall transmission of
the device, see Eq.~(\ref{eq:ballistic}); $c_0
\approx 0.95$ for the 310 nm constriction.  The traces are shifted
horizontally for clarity.
}
	\label{fig:kinks_devices}
\end{figure}

We find regular kink structures in the conductance 
trace of our constriction devices (see, e.g., arrows 
in Fig.~\ref{fig:plateaus_310nm}b). 
%
%===============================================================================================
%===============================================================================================
%
These kinks are well reproducible for different cool-downs 
of the same device (see Figs.~\ref{fig:MultipleCoolDown1}, 
\ref{fig:MultipleCoolDown}), and appear 
in conductance data of several different devices (see  
Fig.~\ref{fig:kinks_devices}).
Analyzing the position of kinks as a function of back-gate voltage 
offers an independent check of the trap state density $n_T$.
%
%===============================================================================================
%===============================================================================================

%===============================================================================================
%===============================================================================================
%
%
%===============================================================================================
\begin{figure*}[!h]
\includegraphics[width=0.93\linewidth]{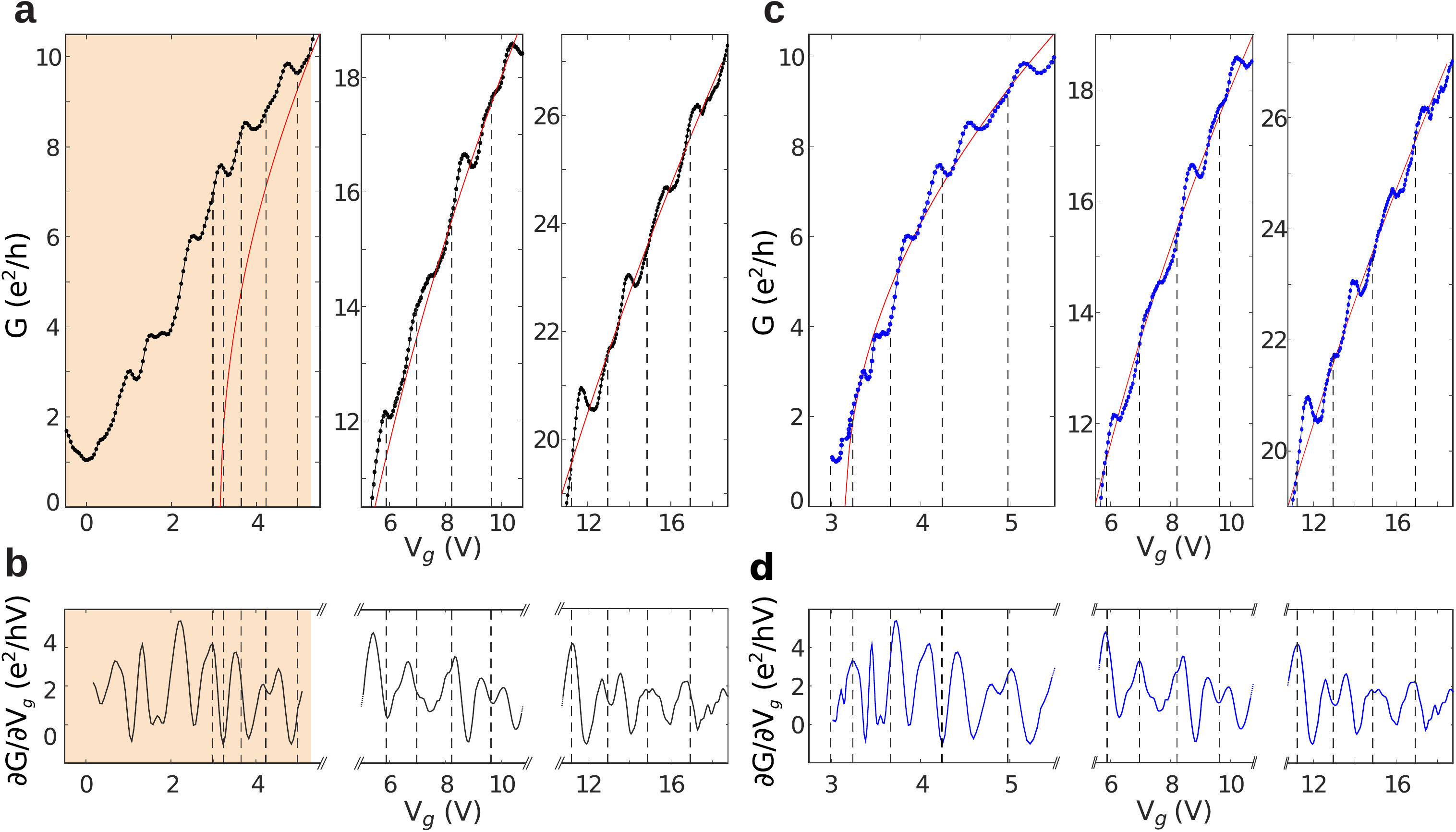}
	\caption[Fig.S3]{\textbf{Back-gate characteristics of the energy subbands of the 230 nm-wide graphene constriction.}
\textbf{(a)} Low-bias four-terminal conductance $G$ as a function of back-gate voltage $V_g$. The theoretical position of the subbands in the $V_{g}$-axis is indicated by vertical dashed lines. Close to the Dirac point (leftmost subpanel) measurements deviate from the ideal Landau model $G \propto \sqrt {V_g}$ shown in red (orange-shaded region).
\textbf{(b)} Derivative plot $\partial G /\partial V_{g}$ of the conductance trace shown in panel (a). The correlation between the expected position of the subbands (vertical dashed lines) and measurements holds only at high carrier densities.
\textbf{(c)} Same as (a) after rescaling of the charge carrier density (Eq.~3). The vertical dashed lines indicating the theoretical position of the subbands matches now the positions of kinks. 
\textbf{(d)} Derivative plot $\partial G /\partial V_{g}$ of the conductance trace in panel (b).
}
	\label{fig:Subbands_Pos}
\end{figure*}

%===============================================================================================
%
In a first order approximation, the band structure of a 
graphene constriction of width $W$ can be described as a 
collection of one-dimensional subbands originating from the quantization 
of the wave vector perpendicular to the transport direction,
\begin{equation}\label{eq:Energy_subbands}
  k_{\perp} = \pm \left|M + \beta \right| \pi / W,
\end{equation}
where $M = 0,\pm 1,\pm 2,\ldots$ is an integer associated with the
subband index (both signs emerge due to the presence of two cones),
and 0 $\leq \left| \beta \right| < 0.5$ is a Maslov index related to
the boundary conditions at the edges (for simplicity we use $\beta=0$,
i.e. a zigzag ribbon).  Within the energy range where the ballistic
model (see red trace in Fig.~\ref{fig:Subbands_Pos}) fits the
conductance trace, the theoretical position of the subbands (marked by
vertical black dashed lines in Fig.~\ref{fig:Subbands_Pos}) for a
$230$ nm-wide graphene constriction ($V^M_{g} = \pi M^2 /\alpha W^2$,
$M=1,2,\ldots$) are in good agreement with the kinks in the
conductance (see Fig.~\ref{fig:Subbands_Pos}a). The agreement
between model and data is also visible in the derivative of the
conductance $\partial G /\partial V_{g}$ (see
Fig.~\ref{fig:Subbands_Pos}b).
Close to the charge neutrality point though, the kink signatures do
not appear to follow the theoretical position of the subbands
(vertical black dashed lines in Fig.~\ref{fig:Subbands_Pos}a,b).
Upon rescaling $k_{\mathrm{F}}$ according to Eq.~(\ref{rescaledkf})
(independently determined from the average transmission), the kinks
are shifted, in good agreement with the quantization model (see
comparison between dashed vertical lines and the position of the kinks
in Fig.~\ref{fig:Subbands_Pos}c,d). In summary, we find that the
rescaling according to Eq.~(\ref{rescaledkf}) will (i) realign
similar, reproducible kink-structures of different cool-downs on the
$k_F$ axis and (ii) shifts the kink positions to fit the simple quantization model of
Eq.~(\ref{eq:Energy_subbands}).

\clearpage

\subsection{Fourier spectroscopy of transmission data}

\begin{figure}[!h]\centering
\includegraphics[draft=false,keepaspectratio=true,clip,%
                   width=0.95\linewidth]%
                   {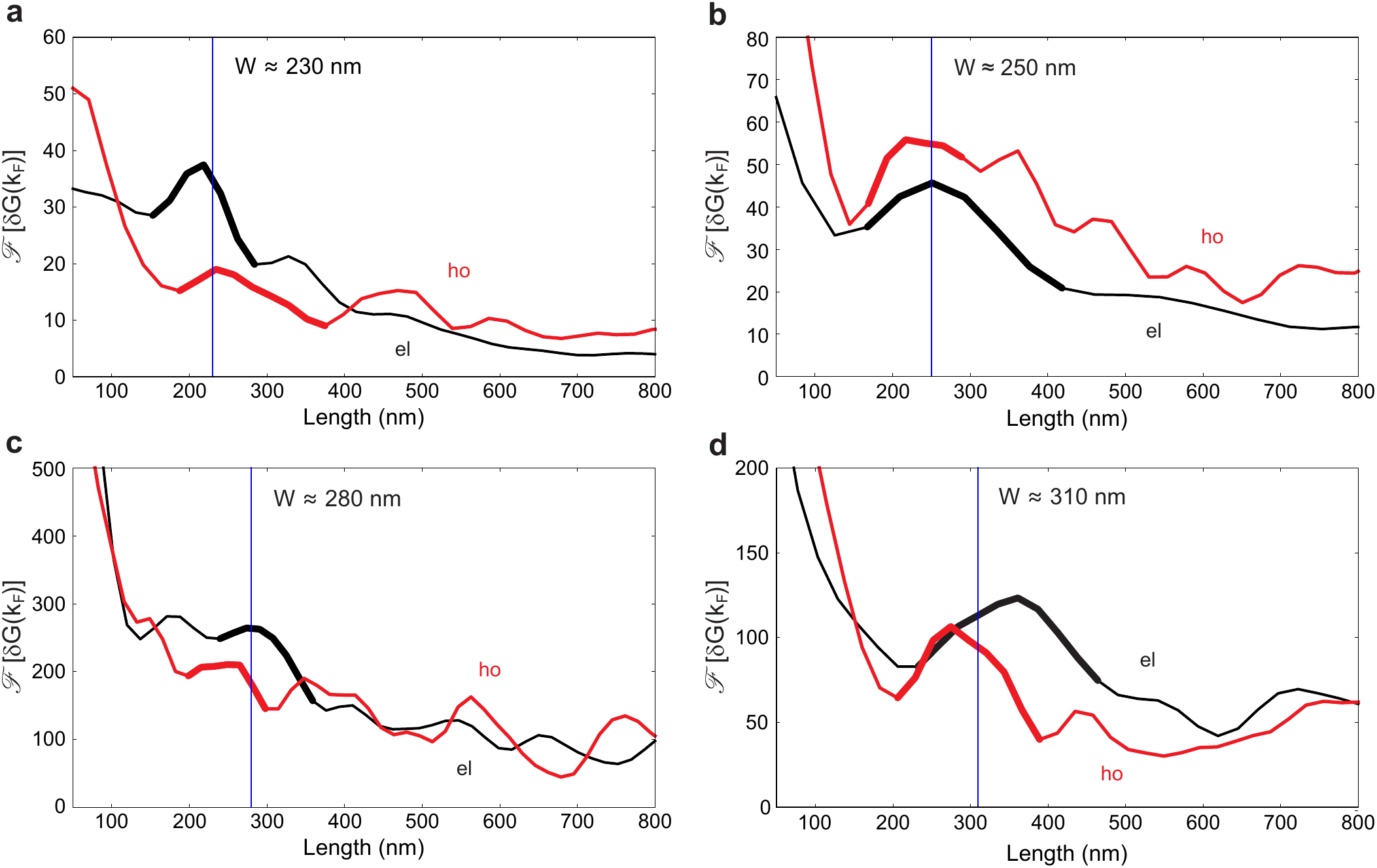}                   
\caption[FIG1]{\textbf{Fourier transform of the conductance.} 
Fourier transform of the electron (black) and hole (red) conductance for the devices of width $230\,$nm (\textbf{a}), $250\,$nm (\textbf{b}), $280\,$nm (\textbf{c}) and $310\,$nm (\textbf{d}). The widths extracted from the Fourier analysis $W_{\mathrm{F}}$ (peaks in $\mathscr{F}[\delta G(k_F)$]) are in good agreement with the widths extracted from SEM images (blue vertical lines). The extracted widths $W_{\mathrm{F}}$ and associated errors bars are shown in Fig.~3f.
}
	\label{fig:FFT}
\end{figure}

Once the conductance is represented as a function of $k_F$, 
the Fourier transform of $\delta G(k_{\mathrm{F}})$ offers alternative 
information on the quantized conductance through the constriction. 
If the regular kinks we identify
in our conductance data, indeed, correspond to size quantization
signatures, we can extract the constriction width from the first peak
of the Fourier transform. Comparison between the first peak in the Fourier
transform of the measured conductance $G(k_\mathrm{F})-G^{(0)}(k_\mathrm{F})$ of four
constriction devices (see Fig.~\ref{fig:FFT} and
Fig.~4 in the main text) to the geometric width $W$ of the
constriction, yields good agreement (see also Fig.~3f of the main
text).

%===============================================================================================

\clearpage

\subsection{Bias spectroscopy}

\begin{figure*}[b!]
  \includegraphics[width=0.88\linewidth]{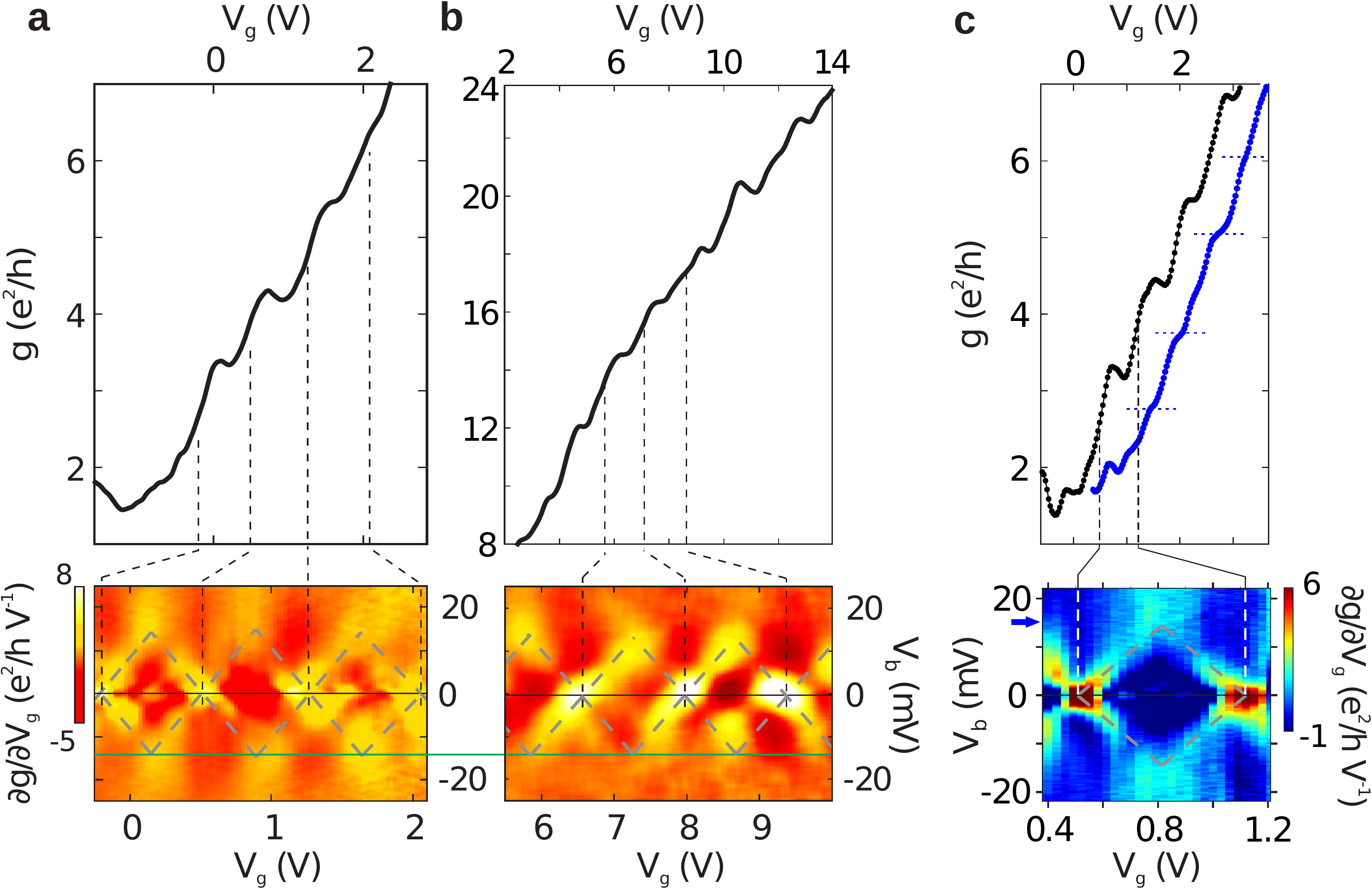}
\caption[Fig.S2]{\textbf{Bias spectroscopy of the 230 nm-wide graphene constriction.}
\textbf{(a)} Differential conductance $g$ (upper panel) and differential transconductance $\partial g/ \partial V_{g}$ (lower panel) as a function of back gate $V_{g}$ and bias $V_{b}$ voltages, measured at $B=0$ T and $T=6$ K. The differential conductance $g$ (top panel) is measured at $V_{\mathrm{b}} = 0$ V in the low carrier density range. The vertical black dashed lines indicate the position of the analyzed subbands.
%The dashed blue horizontal lines indicate the level of conductance 
%of the intermediate plateaus at $V_{\mathrm{b}} = 14$ mV (blue trace).
%The dashed blue horizontal lines indicate the half-conductance values of the intermediate plateaus at $V_{\mathrm{bias}} = 14$ mV.
%
The transconductance $\partial g/ \partial V_{g}$ (bottom color-scaled panel), of the data shown in the upper panel, is measured as a function of an applied bias voltage $V_{\mathrm{b}}$. The kinks are characterized by high values (yellow color) of transconductance. The diamond structures are highlighted by dashed gray diamonds. We extract an average subband spacing $\Delta E \approx 13.5 \,\pm2$ meV (green line). 
\textbf{(b)} Same as panel \textbf{(a)} measured at high carrier densities.
\textbf{(c)} Same as panel \textbf{(a)} for a second cool-down of the same device. The blue trace represents the differential conductance $g$ measured at $V_b = 15\, mV$ (see blue arrow in lower colored panel). The horizontal dashed blue lines highlight the levels of conductance of the intermediate kinks, visible (blue conductance trace) for energies above the subband spacing, e.g. $E \approx 15 \,meV > \Delta E$ (blue arrow in lower colored panel).
}
\label{fig:Bias_Spectro_Diamonds}
\end{figure*}

\begin{figure*}
  \includegraphics[width=0.65\linewidth]{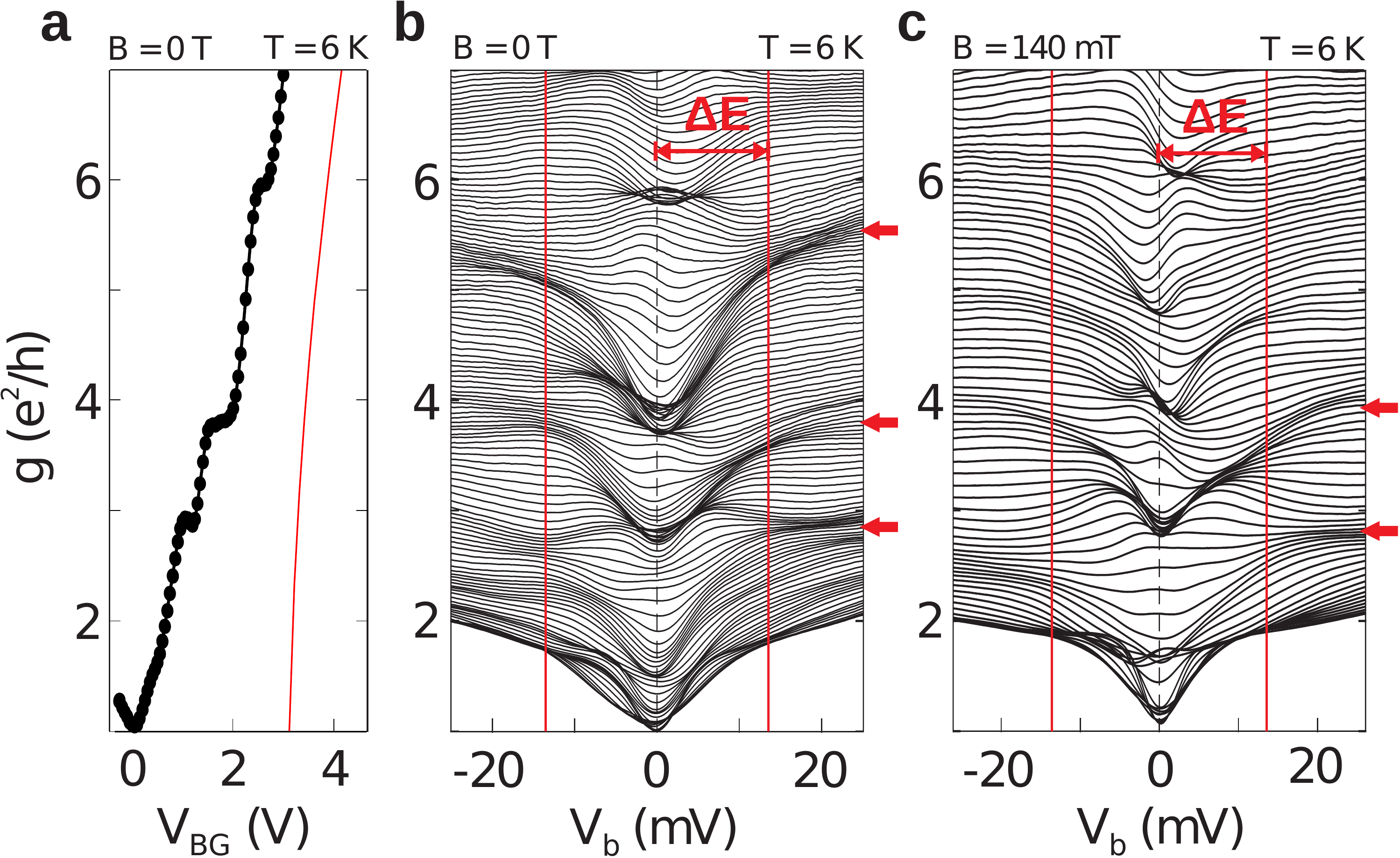}
\caption[Fig.S6]{\textbf{Finite bias spectroscopy of the 230~nm-wide graphene constriction.}
\textbf{(a)} Differential conductance $g$ as a function of back-gate voltage, measured at $V_{b} = 0$ V, $B=0$ T and $T=6$ K. The red solid line shows the ballistic model of conductance, fitted at high carrier densities.
\textbf{(b)} Differential conductance $g$ as a function of source-drain voltage $V_{b}$. The traces are taken at fixed values of back-gate voltage $V_{g}$ from $-0.5$ V (lower trace) to $3.0$ V (upper trace) in steps of $30$ mV. The dense regions correspond to kinks in conductance. The intermediate kinks at high bias voltage are marked by red arrows. The subband spacing ${\Delta E}~\approx~13.5 \pm 3$~meV is highlighted by a vertical red line.
\textbf{(c)} Differential conductance $g$ as a function of source-drain voltage $V_{\mathrm{b}}$ measured at $B=140$~mT. The intermediate kinks at high bias voltage are marked by red arrows. We extract an equal subband spacing as in panel b, $\Delta E \approx 13.5 \pm 3$ meV (vertical red line).
}
  \label{fig:Bias_Spectro}
\end{figure*}

 Using bias spectroscopy we can extract the energy scale associated
 with the regular kink pattern. The differential conductance
 $g\!=\!dI/dV\!=\!I_{SD}/V_{SD}$ (Fig.~4,
 Fig.~\ref{fig:Bias_Spectro_Diamonds} and Fig.~\ref{fig:Bias_Spectro})
 is measured from an AC excitation voltage $V_{AC}\!=\!250\,\mu
 V_{PP}$, using standard Lock-In techniques.  We analyze six diamonds
   associated with kinks at the low- and high-conductance ranges (see
   Fig.~\ref{fig:Bias_Spectro_Diamonds}). Extraction of the energy
   scale from the derivative of the differential conductance (color
   panels) yields $\Delta E = 13.5 \pm 2$ meV leading to $v_F = (1.5
   \pm 0.2) \times 10^6$ m/s.  Variations in the data are due to
   temperature effects, potential variations and uncertainties in
   determining the exact extensions of the diamonds.  All six
   extracted diamonds are taken from energy regions where size
   quantization signatures are clearly visible and reproducible - we
   are thus confident that the sample is in the quantum point contact
   regime for all six diamonds. Note that modifications of the
   gate-lever arm do not affect the bias spectroscopy data since all
   energy scales are extracted from the bias voltage axis ($V_b$),
   which represents a direct energy-scale.

We extract similar values of subband spacing ($\Delta E \approx 13.5 \pm 2$ 
and $13.5 \pm 3\,meV$) in a second (Fig.~4b of the main text 
and Fig.~\ref{fig:Bias_Spectro_Diamonds}c) and a third 
(Fig.~4a of the main text and Fig.~\ref{fig:Bias_Spectro}) 
cool-down of the same device. The value of subband spacing is 
additionally confirmed at finite magnetic field (Fig.~\ref{fig:Bias_Spectro}c). 
We note that, at $B=140$ mT, the quantized subbands are still caused by 
geometric confinement rather than magnetic confinement (i.e., due to the quantum 
Hall effect).

Moreover, half-conductance kinks\cite{Wep13,Har89} are expected to emerge for a bias 
window $e\,V_b$ greater than the subband spacing. Indeed, 
additional kinks at intermediate values of conductance are observed 
(horizontal dashed blue lines in Fig.~\ref{fig:Bias_Spectro_Diamonds}c 
and red arrows in Fig.~\ref{fig:Bias_Spectro}b,c). The observation 
of these intermediate kinks confirms the confinement nature of the 
observed kinks in conductance \cite{Jon73,Kha88,Wep13}.

\begin{figure*}
  \includegraphics[width=0.5\linewidth]{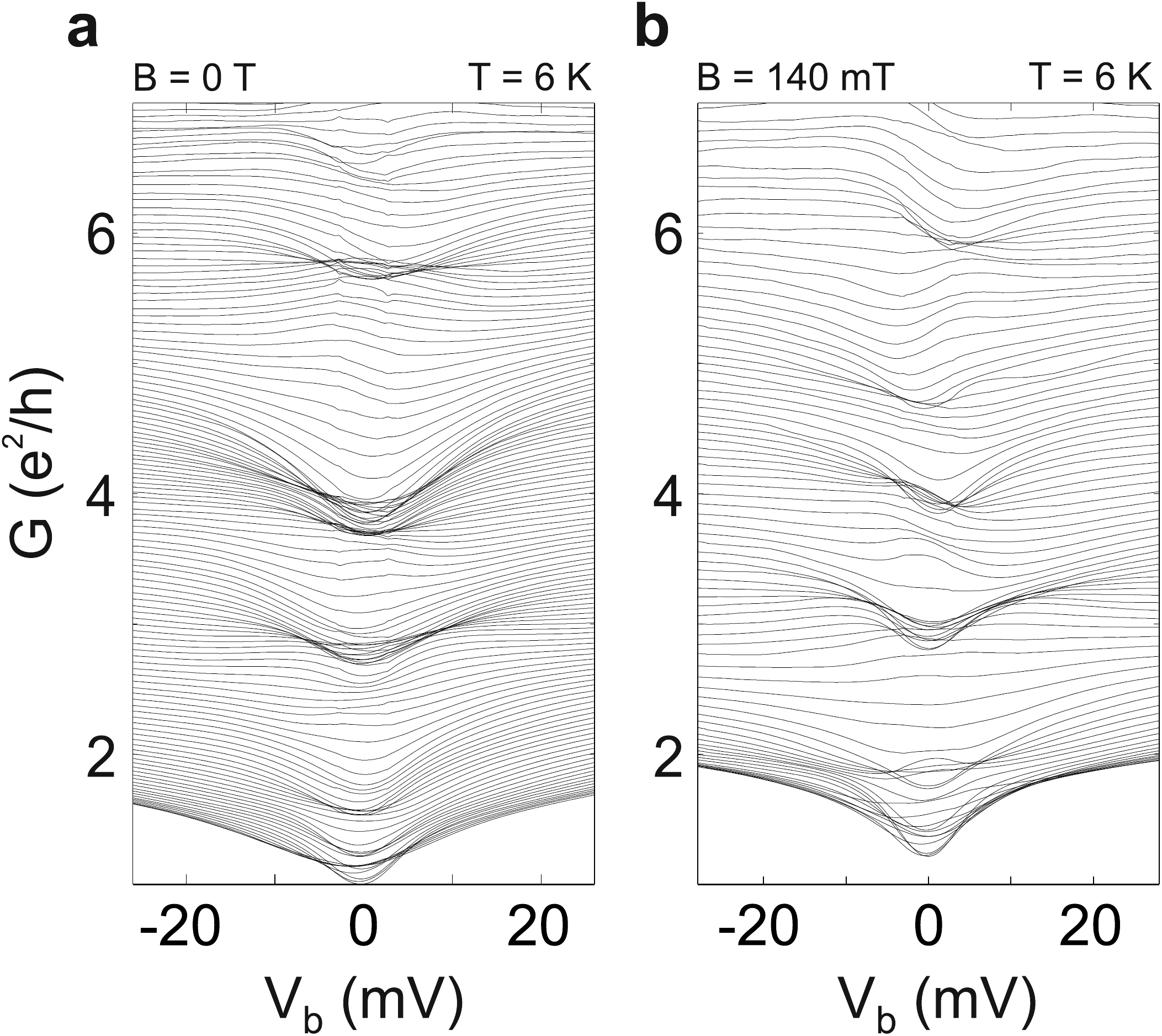}
\caption[Fig.S7]{\textbf{Finite DC bias spectroscopy of the 230~nm-wide graphene constriction.}
\textbf{(a)} DC spectroscopy of the same device as in Figure~\ref{fig:Bias_Spectro}.
\textbf{(b)} Conductance $G$ as a function of DC source-drain voltage $V_{\mathrm{b}}$ measured at $B=140$ mT and $T=6$ K, for the same device as in panel a.
}
  \label{fig:Bias_Spectro_DC}
\end{figure*}

To check against any spurious contribution from the AC measurement
technique, the bias spectroscopy measurements have been repeated 
in a DC configuration (Fig.~\ref{fig:Bias_Spectro_DC}). 
The conductance $G=I/V=I^{DC}/V_{b}$ is obtained from a symmetrically 
applied source-drain DC bias voltage $V_{b}$. Although the resolution 
of the DC conductance $G$ (Fig.~\ref{fig:Bias_Spectro_DC}) 
is not sufficient to extract the subband spacing $\Delta E$, the 
conductance kinks are still visible at identical values of conductance 
as in the AC measurements (Fig.~\ref{fig:Bias_Spectro}).

%===============================================================================================
%===============================================================================================

\clearpage

%===============================================================================================
%===============================================================================================

%===============================================================================================
%===============================================================================================

\clearpage

\subsection{Temperature dependence}

In this note we show additional data on
the temperature dependence of our transport
data highlighting both (i) the high quality 
of our samples and (ii) the energy scale and
stability of the observed kink features.

%===============================================================================================
\begin{figure*}[h]
    \includegraphics[width=0.45\linewidth]{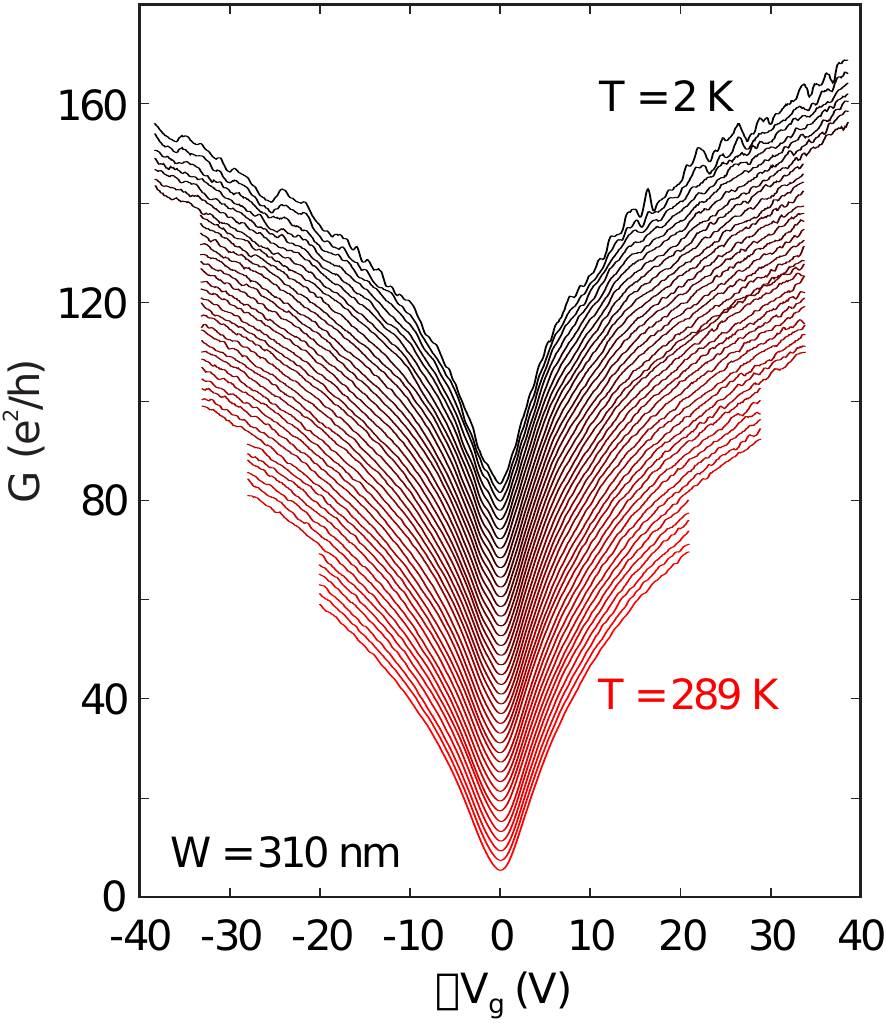}
\caption[Fig.S4]{\textbf{Temperature dependence of the back-gate characteristics for the 310~nm-wide graphene constriction.}
Low-bias back-gate dependent four-terminal conductance $G$ as a function of temperature $T$. The traces are shifted in the conductance axis for clarity. Temperature is recorded from $T=2$ K (black trace) up to room-temperature ($T=289$ K, red trace), in steps of $7$ K.
}
  \label{fig:tem_dep}
\end{figure*}
%===============================================================================================

%===============================================================================================
\begin{figure*}[h]
  \includegraphics[width=0.9\linewidth]{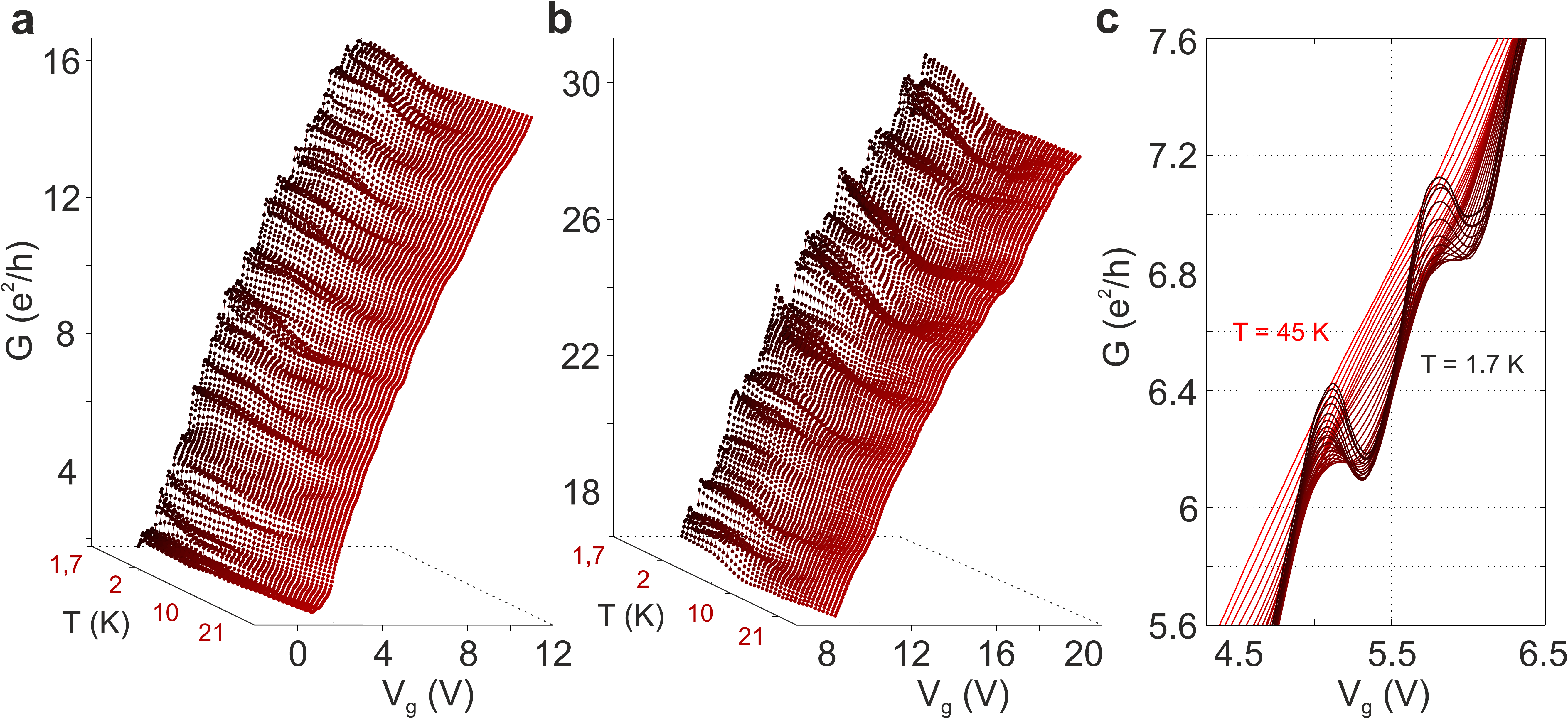}
\caption[Fig.S5]{\textbf{Temperature dependence of the conductance kinks for the 230~nm-wide constriction.}
\textbf{(a)} and \textbf{(b)} Four-terminal conductance $G$ as a function of back gate voltage $V_g$ and temperature $T$, at low (panel a) and high (panel b) carrier densities. Measurements are recorded at temperatures from $T = 2$ K to $T = 24$ K in steps of $0.7$ K.
\textbf{(c)} Zoom-in of the temperature evolution of the shape of the kinks.
% the kink goes away easily as the lpateaus itself -->resoances distord the confinement quantization
}
  \label{fig:tem_dep_2}
\end{figure*}
%===============================================================================================

\newpage

\subsection{Evolution of size quantization with magnetic field}

We provide an additional data set for the magnetic-field evolution of
the size quantization signatures from the $280\,$nm-wide graphene
constriction in Fig. S15. We find the same
transition from size-quantization signatures, at low magnetic fields, 
to the Landau level regime, at high magnetic fields, as in the sample
discussed in the main text (see Fig.~5 of main manuscript).

%===============================================================================================
%\begin{figure*}[h]
%  \includegraphics[width=1.0\linewidth]{Fig05Suppl_v0_3_1.pdf}
%\caption[Fig.S15]{\textbf{Magnetic-field dependence of the size quantization for the \textcolor[rgb]{1,0,0}{280} nm-wide graphene constriction.} 
%%
%\textbf{a}, Landau level fan from the derivative of the longitudinal conductance. The first Landau levels emerge at magnetic field on the order of $1$ T. 
%%
%\textbf{b} and \textbf{c}, Derivative of the longitudinal conductance from a high resolution measurement for magnetic field $B \leq 1$ T in the low%-carrier density range on the hole side and electron side respectively.  In panels a, b and c, the black dashed line denotes the boundary above which the %magnetic field quantization of Landau level $m$ dominates over size quantization, i.e. when $2\sqrt{2m}l_B < W$. 
%%
%\textbf{d} Evolution of the conductance versus charge carrier density traces as a function of magnetic field $B$. The field step size between each trace %is $8$ mT. This data set was measured at $T = 1.7$~K. 
%%
%}
%\label{fig:magnetodep}
%\end{figure}
%===============================================================================================

%===============================================================================================
\begin{figure*}[h]
  \includegraphics[width=1.0\linewidth]{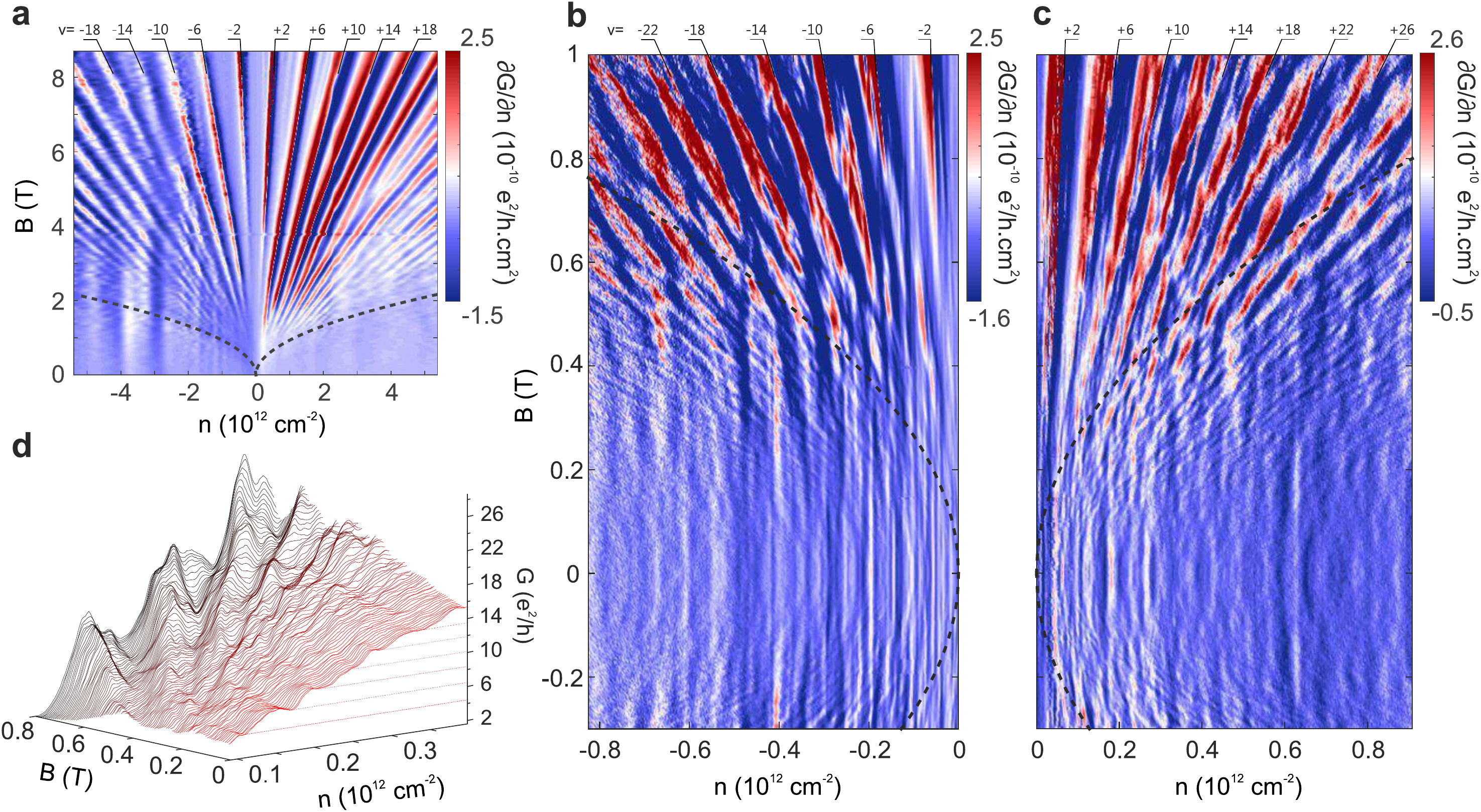}
\caption[Fig.S15]{\textbf{Magnetic-field dependence of the size quantization for the 280~nm-wide graphene constriction.}
\textbf{(a)} Landau level fan of the 280 nm-wide graphene constriction.
\textbf{(b)} and \textbf{(c)} High resolution double derivative plots, measured at low magnetic fields $B \leq 1$ T, in the low-carrier density range for the hole- and electron-regimes, respectively. In panels a, b and c the black dashed line denotes the boundary above which the magnetic field quantization of Landau level $m$ dominates over size quantization, i.e. when $2 \sqrt{2 m} l_B < W$.
\textbf{(d)} Evolution of the conductance traces as a function of charge carrier density $n$ and magnetic field $B$ . The $B$-field step size between traces is $8$ mT. The data was measured at $T = 1.7$~K.
}
  \label{fig:magnetodep}
\end{figure*}
%===============================================================================================

\subsection{Theoretical treatment}
\label{sec:theory}

We use a third nearest neighbor tight-binding approach to simulate the
constriction. We pattern the device edge using the experimental
geometry determined from SEM, and a correlated random fluctuation to
simulate microscopic roughness. We rescale our device by a factor of
four compared to the experiment, to arrive at a numerically feasible
system size.  Such a rescaling by a factor of four ensures that all
relevant length scales of the problem (e.g., device geometry, Fermi
wavelength, magnetic length and correlation length of the edge
roughness) are still much larger than the discretization length of the
numerical graphene lattice, allowing to extrapolate simulation data to
the experimental result~\cite{liu15}.  We use a correlation length of
5 nm and an average disorder amplitude of 13 nm. We determine the
Green's function, $\mathcal G(r,r')$, of the device using the modular
recursive Green's function method \cite{Rotter06, LibischNJP}. The
local density of states, $\rho(r,E)$, is given by
$\rho(r,E)\propto \mathrm{Im}[ \mathcal G(r,r;E)]$. Calculations were
performed on the Vienna Scientific Cluster 3. To determine the
transport properties of the device, we attach two leads of width $D$
on each side of the experimental contact regions, and calculate the
total transmission. To avoid residual effects due to the fixed lead
width used in the computation, we average over five different randomly
chosen lead widths $D \in [60,80]$ nm.  

To determine the evolution of subbands in a constriction of width $W$
with magnetic field, we calculate the band structure of a perfect
zigzag graphene nanoribbon of width $W$ as a function of magnetic
field. We include the magnetic field via a Peierls phase factor. The
subband positions are extracted from the minima of each band in the
bandstructure of the ribbon.

\newpage

%%%%%%%%%%%%%%%%%%%%%%%%%%%%%%%%%%%%%%%%%%%%%%%%%%%%%%%%%%%%%%%%%%%%%%%
%\bibliographystyle{naturemag}
%\bibliography{Bib_NaturePhys}

\begin{thebibliography}{10}

\bibitem{You09}								%% #1
Young,\,A.F. \& Kim,\,P. 
Quantum interference and Klein tunnelling in graphene heterojunctions.
\textsl{Nature Phys.} \textbf{5}, 222 (2009).

\bibitem{Two06}								%% #2
Tworzydlo,\,J., et al.
%Trauzettel,\,B., Titov,\,M., Rycerz,\,A. \& Beenakker,\,C.W.J.
Sub-Poissonian Shot Noise in Graphene.
\textsl{Phys. Rev. Lett.} \textbf{96}, 246802 (2006).

\bibitem{Nov05}								%% #3
Novoselov,\,K.S., et al.
%Geim,\,A.K., Morozov,\,S.V., Jiang,\,D., Katsnelson,\,M.I., Grigorieva,\,I.V., Dubonos,\,S.V. \& Firsov,\,A.A.
Two-dimensional gas of massless Dirac fermions in graphene.
\textsl{Nature} \textbf{438}, 197 (2005).

\bibitem{Zha05}								%% #4
Zhang,\,Y., Tan,\,Y-W. Stormer,\,H.L. \& Kim,\,P.
Experimental observation of the quantum Hall effect and Berry's phase in graphene. 
\textsl{Nature} \textbf{438}, 201 (2005).

\bibitem{Du09}								%% #5
Du,\,X., Skachko,\,I., Duerr,\,F., Luican,\,A., \& Andrei,\,E.Y.
Fractional quantum Hall effect and insulating phase of Dirac electrons in graphene. 
\textsl{Nature} \textbf{462}, 192 (2009).

\bibitem{Bol09}								%% #6
Bolotin,\,K.I., et al.
%Ghahari,\,F., Shulman,\,M.D., Stormer,\,H.L. \& Kim,\,P.
Observation of the fractional quantum Hall effect in graphene.
\textsl{Nature} \textbf{462}, 196 (2009).

\bibitem{Dean10}							%% #7
Dean,\,C.R., et al.
%Young,\,A.F., Meric,\,I., Lee,\,C., Wang,\,L., Sorgenfrei,\,S., Watanabe,\,K., Taniguchi,\,T. Kim,\,P., Shepard,\,K.L., \& Hone,\,J.
Boron nitride substrates for high-quality graphene electronics.
\textsl{Nature Nano.} {\bf 5}, 722 (2010).

\bibitem{Wan13}								%% #24
Wang,\,L., et al.
%I.Meric, P.Y.Huang, Q.Gao, Y.Gao, H.Tran, T.Taniguchi, K.Watanabe, L.M.Campos, D.A.Muller, J.Guo, P.Kim, J.Hone, K.L.Shepard, \& C.R.Dean, 
One-Dimensional Electrical Contact to a Two-Dimensional Material. 
\textsl{Science} \textbf{342}, (6158) 614-167 (2013).

\bibitem{Lin08}								%% #9
Lin,\,Y.M., Perebeinos,\,V., Chen,\,Z., \& Avouris,\,P.
Electrical observation of subband formation in graphene nanoribbons. 
\textsl{Phys. Rev. B} \textbf{78}, 161409R (2008).

\bibitem{wang11}	
Wang,\,X., et al.
%Yijian Ouyang,	Liying Jiao,	Hailiang Wang,	Liming Xie,	Justin Wu,	Jing Guo	& Hongjie Dai	
Graphene nanoribbons with smooth edges behave as quantum wires.
\textsl{Nature Nanotechnology} \textbf{6}, 563–567 (2011).

\bibitem{Tom11}								%% #10
Tombros,\,N., et al.
%Veligura,\,A., Junesch,\,J., Guimaraes,\,M.H.D., Vera-Marun,\,I.J., Jonkman,\,H.T. \& van Wees,\,B.J. 
Quantized conductance of a suspended graphene nanoconstriction. 
\textsl{Nature Physics} \textbf{7}, 697-700 (2011).

\bibitem{Ter11}								%% #11
Terr\'es,\,B., et al.
%Dauber,\,J., Volk,\,C., Trellenkamp,\,S., Wichmann,\,U. \& Stampfer,\,C. 
Disorder induced Coulomb gaps in graphene constrictions with different aspect ratios. 
\textsl{Appl. Phys. Lett.} \textbf{98}, 032109 (2011).

\bibitem{Sarma11}							%% #12
Das Sarma,\,S., Adam,\,S., Hwang,\,E.H. \& Rossi,\,E.
Electronic Transport in 2D Graphene.  
\textsl{Rev. Mod. Phys.} {\bf 83}, 407 (2011).

\bibitem{Danneau08}						%% #13
Danneau,\,R. et al.
%F. Wu, M. F. Craciun, S. Russo, M. Y. Tomi, J. Salmilehto, A. F. Morpurgo, and P. J. Hakonen
Shot Noise in Ballistic Graphene.
\textsl{Phys. Rev. Lett.} {\bf 100}, 196802 (2008).

\bibitem{Borunda13}						%% #14
Borunda,\,M.F., Hennig,\,H. \& Heller,\,E.J.
Ballistic versus diffusive transport in graphene. 
\textsl{Phys. Rev. B} {\bf 88}, 125415 (2013).

\bibitem{Mas12}								%% #15
Masubuchi,\,S., et al.
%Iguchi,\,K., Yamaguchi,\,T., Onuki,\,M., Arai,\,M., Watanabe,\,K., Taniguchi,\,T. \& Machida,\,T. 
Boundary Scattering in Ballistic Graphene. 
\textsl{Phys. Rev. Lett.} \textbf{109}, 036601 (2012).

\bibitem{Baringhaus14}				%% #16
Baringhaus,\,J., et al.
%M. Ruan, F. Edler, A. Tejeda, M. Sicot, A. Taleb-Ibrahimi, A.-P. Li, Z. Jiang, E. H. Conrad, C. Berger, C. Tegenkamp, and W. A. de Heer, 
Exceptional ballistic transport in epitaxial graphene nanoribbons.
\textsl{Nature} {\bf 506}, 349 (2014)

\bibitem{Magda14}							%% #17
Magda,\,G.Z., et al.
%Jin,\,X., Hagym\'asi,\,I., Vancs\'o,\,P., Osv\'ath,\,Z., Nemes-Incze,\,P., Hwang,\,C., Bir\'o,\,L.P. \& Tapaszt,\,L.
Room-temperature magnetic order on zigzag edges of narrow graphene nanoribbons.
\textsl{Nature} \textbf{514}, 608 (2014)

\bibitem{Titov}
M.~Titov \& C. W. J. Beenakker
Josephson effect in ballistic graphene.
\textsl{Phys. Rev. B.} \textbf{74}, 041401(R) (2006).

\bibitem{Plotnik14}						%% #18
Plotnik,\,Y., et al.
%Rechtsman,\,M.C., Song,\,D., Heinrich,\,M., Zeuner,\,J.M., Nolte,\,S., Lumer,\,Y., Malkova,\,N., Xu,\,J., Szameit,\,A., Chen,\,Z. \& Segev,\,M.
Observation of unconventional edge states in ‘photonic graphene’.
\textsl{Nature Mat.} \textbf{13}, 57-62 (2014)

\bibitem{Yang08}							%% #19
Yang,\,L., Cohen,\,M.L. \& Louie,\,S.G.
Magnetic Edge-State Excitons in Zigzag Graphene Nanoribbons.
\textsl{Phys. Rev. Lett.} \textbf{101}, 186401 (2008).

\bibitem{Young14}							%% #20
Young,\,A.F., et al.
%Sanchez-Yamagishi,\,J.D., Hunt,\,B., Choi,\,S.H., Watanabe,\,K., Taniguchi,\,T., Ashoori,\,R.C. \& Jarillo-Herrero,\,P.
Tunable symmetry breaking and helical edge transport in a graphene quantum spin Hall state.
\textsl{Nature} \textbf{505}, 528 (2014).

\bibitem{Ostaay11}						%% #21
Van Ostaay,\,J.A.M., et al.
%Akhmerov,\,A.R., Beenakker,\,C.W.J. \& Wimmer,\,M.
Dirac boundary condition at the reconstructed zigzag edge of graphene.
\textsl{Phys. Rev. B} {\textbf 84}, 195434 (2011).

\bibitem{YZhang05}						%% #22
Zhang,\,Y., Tan,\,Y.-W., Stormer,\,H.L., \& Kim,\,P.
Experimental observation of the quantum Hall effect and Berry's phase in graphene. 
\textsl{Nature} {\bf 438}, 201 (2005).

\bibitem{Novoselov07}					%% #23
Novoselov,\,K.S., et al.
%Z. Jiang, Y. Zhang, S. V. Morozov, H. L. Stormer, U. Zeitler, J. C. Maan, G. S. Boebinger, P. Kim, and  A. K. Geim, 
Room-Temperature Quantum Hall Effect in Graphene.
\textsl{Science} {\bf 315}, 1379 (2007).

\bibitem{Reiter14}						%% #25
Reiter,\,R., et al.
%Derra,\,U., Birner,\,S., Terr\'es,\,B., Libisch,\,F., Burgd\"orfer,\,J. \& Stampfer,\,C.
Negative quantum capacitance in graphene nanoribbons with lateral gates.
\textsl{Phys. Rev. B} {\bf 89}, 115406 (2014).

\bibitem{Ilani06}							%% #26
Ilani,\,S., et al.
%L. A K. Donev, M. Kindermann, and P. L. McEuen, 
Measurement of the quantum capacitance of interacting electrons in carbon nanotubes.
\textsl{Nature Physics} {\bf 2}, 687 (2006).

\bibitem{Fang07}							%% #27
Fang,\,T., et al.
% A. Konar, H. Xing, and D. Jena, 
Carrier statistics and quantum capacitance of graphene sheets and ribbons.
\textsl{App. Phys. Lett.} {\bf 91}, 092109 (2007).

\bibitem{LeRoySTM}
A. Deshpande, W. Bao, Z. Zhao, C. N. Lau, \& B. J. LeRoy
Imaging charge density fluctuations in graphene using Coulomb blockade spectroscopy
\textsl{Phys. Rev. B} {\bf 83}, 155409 (2011).

\bibitem{Bischoff14} 					%% #28
Bischoff,\,D., et al.
%F.~Libisch, J.~Burgd\"orfer, T.~Ihn, and K.~Ensslin,
Characterizing wave functions in graphene nanodevices: Electronic transport through ultrashort graphene constrictions on a boron nitride substrate.
\textsl{Phys.~Rev.~B} \textbf{90}, 115405 (2014).

\bibitem{LibischNJP} 					%% #29
Libisch,\,F., Rotter,\,S., \& Burgd\"orfer,\,J.
Coherent transport through graphene nanoribbons in the presence of edge disorder.
\textsl{New Journal of Physics} \textbf{14}, 123006 (2012).

\bibitem{liu15}
Liu,\,M.-H. et al. \textsl{Phys. Rev. Lett.} \textbf{114}, 036601 (2015).

\bibitem{Per06}								%% #30
Peres,\,N.M.R., et al.
%Castro Neto,\,A.H., \& Guinea,\,F.
Conductance quantization in mesoscopic graphene.
\textsl{Phys. Rev. B} \textbf{73}, 195411 (2006).

\bibitem{Muc09}								%% #31
Mucciolo,\,E.R., et al.
%Castro Neto,\,A.H. \& Lewenkopf,\,C.H. 
Conductance quantization and transport gaps in disordered graphene ribbons.
\textsl{Phys. Rev. B} \textbf{79}, 075407 (2009).

\bibitem{Ihnatsenka12}								%% #31
Ihnatsenka,\,S. \& Kirczenow,\,G
%Castro Neto,\,A.H. \& Lewenkopf,\,C.H. 
Conductance quantization in graphene nanoconstrictions with mesoscopically smooth but atomically stepped boundaries.
\textsl{Phys. Rev. B} \textbf{85}, 121407(R) (2012).

\bibitem{vanWeesConductance}	%% #32
Van Wees,\,B.J., et al.
%H.van Houten, C.W.J.Beenakker, J.G.Williamson, L.P.Kouwenhoven, D.van der Marel, \& C.T.Foxon 
Quantized conductance of point contacts in a two-dimensional electron gas.
\textsl{Phys. Rev. Lett.} \textbf{60}, 848 (1988).

\bibitem{wep13}								%% #33
Van Weperen,\,I., et al.
%S.R. Plissard, E.P.A.M. Bakkers, S.M. Frolov, \& L.P. Kouwenhoven
Quantized Conductance in an InSb Nanowire.
\textsl{Nano Letters} {\bf 13}, 387 (2013).

\bibitem{Elias11}							%% #34 
Elias,\,D.C., et al.
%R. V. Gorbachev, A. S. Mayorov, S. V. Morozov, A. A. Zhukov, P. Blake, çL. A. Ponomarenko, I. V. Grigorieva, K. S. Novoselov, F. Guinea, and A. K. Geim, 
Dirac cones reshaped by interaction effects in suspended graphene.
\textsl{Nature Physics} {\bf 7}, 701 (2011)

\bibitem{gui12}								%% #35
Guimaraes,\,M.H.D.,  et al.
%O. Shevtsov, X. Waintal, \& B. J. van Wees
From quantum confinement to quantum Hall effect in graphene nanostructures.
\textsl{Phys. Rev. B} {\bf 85}, 075424 (2012).

\bibitem{Rotter06} 						%% #36
Rotter,\,S., et al.
%J.Z.Tang, L.Wirtz, J.Trost, \& J.Burgd\"{o}rfer,
Modular recursive Green’s function method for ballistic quantum transport. 
\textsl{Phys. Rev. B} {\bf 62,} 1950 (2000).

\end{thebibliography}

\begin{thebibliography}{10}
  %\bibitem{buttiker:85} 
   % M. B\"uttiker, Y. Imry, R. Landauer, and S. Pinhas, Phys. Rev. B
 %\textbf{31}, 6207 (1985).
 %\bibitem{imry}
%Y. Imry,
% in \emph{Directions in Condensed Matter Physics}, edited by G. Grinstein and
%  G. Mazenko,
% vol. 1 (World Scientific, Singapore, 1986).

%\bibitem{Mar13}
%I. J. Vera-Marun, P. J. Zomer, A. Veligura, M. H. D. Guimarães, L. Visser, N. Tombros, H. J. van Elferen, U. Zeitler and B. J. van Wees,
%Appl. Phys. Lett. 102, 013106 (2013)

\bibitem{Couto14}
N. J. G. Couto, et al.,
Phys. Rev. X 4, 041019 (2014).

\bibitem{Woe16}
A. Woessner, et al.,
arXiv:1508.07864 (2016).

\bibitem{Wep13}
I.~van Weperen, et al., 
Nano Lett., 13, 387-391 (2013).

\bibitem{Har89}
L.~P. Kouwenhoven, et al.,
Phys. Rev. B {\bf 39}, 8040 (1989).

\bibitem{Jon73}
N.~K. Patel, et al.,
Phys. Rev. B {\bf 44}, 10973 (1991).

\bibitem{Kha88}
L.~I. Glazman, et al.,
JETP Lett. {\bf 48}, 591-595 (1988).

\bibitem{liu15}
Liu,\,M.-H. et al., 
Phys. Rev. Lett. \textbf{114}, 036601 (2015).

\bibitem{LibischNJP} 
F.~Libisch, S.~Rotter, and J.~Burgd\"orfer, 
\textit{New Journal of Physics} {\textbf 14}, 123006 (2012).

\bibitem{Rotter06} S.~Rotter, J.~Z.~Tang, L.~Wirtz, J.~Trost, and J.~Burgd\"{o}rfer, 
{\it Phys.~Rev.~B} {\bf 62,} 1950-1960 (2000).

\end{thebibliography}

%\end{document}

\end{document}